\documentclass[aps,prx,onecolumn,amsmath,amssymb,superscriptaddress,floatfix,longbibliography]{revtex4-1}

\usepackage{graphicx}
\usepackage{multirow}
\usepackage{xcolor}
\usepackage{bm}
\usepackage{amsfonts,amssymb,amsmath}
\usepackage{graphicx,dcolumn,bm,color,braket,slashed}
\usepackage{times} 
\usepackage{comment}
\usepackage{array}
\usepackage{textcomp}

\renewcommand{\v}[1]{{\bf #1}}

\begin{document}

\title{Bulk-interface correspondence from quantum distance in flat band systems}

\author{Chang-geun Oh}
\email{cg.oh.0404@gmail.com}
\affiliation{Research Institute for Basic Science, Ajou University, Suwon 16499, Korea}

\author{Doohee Cho}
\affiliation{Department of Physics, Yonsei University, Seoul, 03722, Korea}

\author{Se Young Park}
\affiliation{Department of Physics and Origin of Matter and Evolution of Galaxies (OMEG) Institute, Soongsil University, Seoul 06978, Korea}

\author{Jun-Won Rhim}
\email{jwrhim@ajou.ac.kr}
\affiliation{Department of Physics, Ajou University, Suwon 16499, Korea}
\affiliation{Research Center for Novel Epitaxial Quantum Architectures, Department of Physics, Seoul National University, Seoul, 08826, Korea}

\begin{abstract}
\textbf{Abstract} The bulk-boundary correspondence is an integral feature of topological analysis and the existence of boundary or interface modes offers direct insight into the topological structure of the Bloch wave function.
While only the topology of the wave function has been considered relevant to boundary modes, we demonstrate that another geometric quantity, the so-called quantum distance, can also host a bulk-interface correspondence.
We consider a generic class of two-dimensional flat band systems, where the flat band has a parabolic band-crossing with another dispersive band.
While such flat bands are known to be topologically trivial, we show that the nonzero maximum quantum distance between the eigenstates of the flat band around the touching point guarantees the existence of boundary modes at the interfaces between two domains with different chemical potentials or different maximum quantum distance.
Moreover, the maximum quantum distance can predict even the explicit form of the dispersion relation and decay length of the interface modes.
\end{abstract}

\maketitle

\textbf{Introduction} Geometric properties of Bloch wave functions, such as the Berry curvature and Berry connection, have been the central theme of modern solid-state physics since the discovery of the topological insulator~\cite{kane2005quantum,kane2005z,bernevig2006quantum,bernevig2006quantum_science,fu2007topological}.
Using such geometric quantities, various bulk numbers called the topological invariants can be defined, which encode the topological nature of solid-state systems and enable us to classify phases more delicately beyond the order-parameter-based classifications~\cite{thouless1982quantized,moore2007topological,schnyder2008classification,ando2015topological}.
Representative topological invariants include the Chern number of Chern insulators, Z$_2$ index of topological insulators~\cite{thouless1982quantized,haldane1988model}, mirror Chern number of topological crystalline insulators~\cite{fu2007topological,teo2008surface}, monopole charge of Weyl semimetals~\cite{wan2011topological}, and Zak phase of one-dimensional mirror-symmetric insulators~\cite{zak1989berry,atala2013direct,rhim2017bulk}.

The most direct physical manifestation of these abstract topological orders is the existence of boundary modes that are robust against external perturbations, respecting the symmetries of the given system.
This feature is called the bulk-boundary correspondence, which has played the most central role in the topological analysis of solids~\cite{hatsugai1993chern,kitaev2009periodic,fukui2012bulk,mong2011edge,prodan2016bulk,rhim2018unified}.
As a result, even if a solid is deemed to be an insulator according to the bulk band structure, it can be a metal owing to the boundary modes if the bulk has a nontrivial topological order.
In other words, the bulk-boundary correspondence offers direct access to the bulk topology and helps us figure out the most fundamental material character, i.e., metallic or insulating.

In the perspective of geometry, there is another intriguing quantity called the Hilbert-Schmidt quantum distance~\cite{buvzek1996quantum,dodonov2000hilbert,wilczek1989geometric}, which is defined as
\begin{eqnarray}
        d_{\text{HS}}^2 (\bold{k} , \bold{k'}) = 1 - |\braket{\psi_{\bold{k}}|\psi_{\bold{k'}}}|^2, 
\end{eqnarray}
where $\psi_{\bold{k}}$ is an eigenstate of the Bloch Hamiltonian matrix with crystal momentum $\bold{k}$.
This quantity, also called quantum distance for short, takes the role of distance in the geometric description of Bloch wave functions, while the Berry curvature plays the curvature part.
Indicating how close two states are, the quantum distance is not quantized and varies continuously between 0 and 1.
We show that the quantum distance can be used to predict the existence and band structures of interface modes in two-dimensional flat band systems.
Although the quantum distance is not entirely independent of topological notions \cite{provost1980riemannian,ozawa2021relations,mera2022relating}, there has been no bulk-interface correspondence directly employing this kind of geometric quantity.

\begin{figure}[t]
\includegraphics[width=85mm]{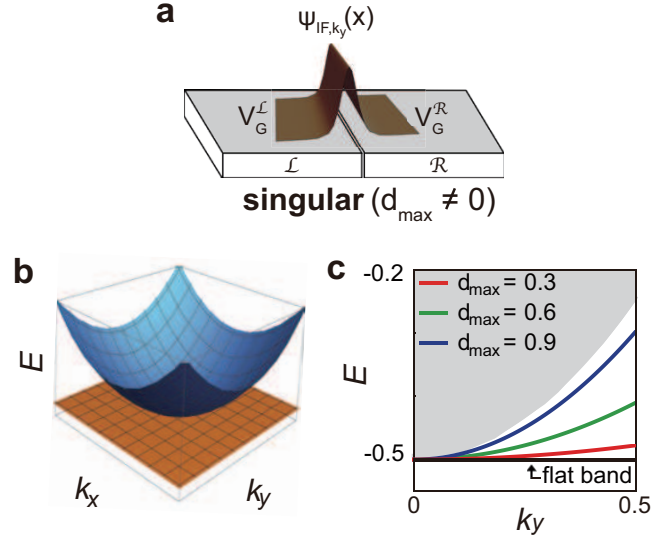} 
\caption{\label{fig1}
\textbf{Singular flat band and interface modes characterized by the quantum distance.}
\textbf{a}, Schematics of the interface mode. When the flat band is singular, namely the maximum quantum distance $d_\mathrm{max}$ of the flat band is nonzero, an interface mode $\Psi_{\mathrm{IF},k_y}$ with a crystal momentum $k_y$ appears between two areas (denoted by $\mathcal{L}$ and $\mathcal{R}$) with different onsite potentials  $V^\mathcal{L}_\mathrm{G}$ and $V^\mathcal{R}_\mathrm{G}$. 
\textbf{b}, A flat band with a parabolic band-crossing. \textbf{c}, The band spectra of interface modes for various values of $d_\mathrm{max}$. The gray region represents the bulk parabolic band. Here, the effective mass tensor of the parabolic band is given by $m^{-1}_{xx}=m^{-1}_{yy}=2$, $m^{-1}_{xy}=0$, and the onsite potentials are set to be $V^\mathcal{L}_\mathrm{G}=0$ and $V^\mathcal{R}_\mathrm{G}=-0.5$.
}\label{fig:main}
\end{figure}

A flat band is a perfectly dispersionless band and has received considerable attention from the perspective of many-body physics because the electron-electron interaction strength is the most dominant energy scale in such a band\cite{morell2010flat,cao2018unconventional,balents2020superconductivity,peri2021fragile,liu2021spectroscopy,sharpe2019emergent,wu2021chern,saito2021hofstadter,hase2018possibility,aoki1996hofstadter,ramachandran2017chiral,regnault2011fractional}.
Recently, a class of flat bands, called singular flat bands (SFBs), has attracted much attention because of their intriguing geometric effects~\cite{rhim2019classification,ma2020direct}.
The Bloch wave function of the SFB has a singularity originating from a band-crossing with another band~\cite{rhim2021singular}.
The band-crossing point is characterized by a geometric quantity called the maximum quantum distance ($d_\mathrm{max}$) if the band-crossing is quadratic type [see Fig.~\ref{fig:main}b].
The maximum quantum distance is the maximum value of quantum distance among all the possible pairs of Bloch eigenstates around the band-crossing point. 
A finite value of $d_\mathrm{max}$ guarantees the existence of topological line-shaped eigenmodes called the non-contractible loop states~\cite{rhim2019classification} and manifests itself as an anomalous Landau level structure, where Landau levels corresponding to the flat band are spread into the nearby band gap, and the magnitude of the spreading is determined by $d_\mathrm{max}$~\cite{rhim2020quantum}.

This work shows that if we introduce an interface in the middle of a singular flat band system with nonzero $d_\mathrm{max}$, localized modes are guaranteed to exist around the interface whose energy dispersion is located between the flat and parabolic bands [see Fig.~\ref{fig:main}a and c].
Here, we consider two types of interfaces.
One is created by applying a potential difference between two 2D flat band system regions, and the other is obtained by making a junction of two different flat band models.
Moreover, while the topological invariants predict only the number of the boundary modes in the gap, we find that $d_\mathrm{max}$ also determines even the effective mass of the band dispersion and the localization length of the interface mode.
Our results show that geometric quantities other than topological notions also can be used for the bulk-interface correspondence.

We consider the most general form of the 2D continuum Hamiltonian describing a flat band with a parabolic band touching [see Fig.~\ref{fig:main}b] given by
    \begin{eqnarray}
        \mathcal{H}_{\mathrm{fb}}(\bold{k}) = \sum_{\alpha } f_\alpha (\bold{k}) \sigma_\alpha , \label{eq:Ham}
    \end{eqnarray}
where $\sigma_\alpha$ represents an identity ($\alpha=0$) and Pauli matrices ($\alpha = x,y,z$)~\cite{rhim2019classification}.
See Supplementary Note 1 for details.
Here, $f_\alpha (\bold{k})$ is a real quadratic function: $f_{x} (\bold{k}) = t_6 k_y^2,~f_{y} (\bold{k}) = t_4 k_x k_y + t_5 k_y^2,~f_{z} (\bold{k}) = t_1 k_x^2 + t_2 k_x k_y + t_3 k_y^2 $, and $f_{0} (\bold{k}) = b_1 k_x^2 + b_2 k_x k_y + b_3 k_y^2$, where $t_i$ and $b_i$ are real-valued band parameters.
Any kind of quadratic Hamiltonian can be unitarily transformed into the above.
Due to the flat band condition, only four out of nine band parameters are independent, and can be chosen as $P_0 = \{ t_1,t_2,t_3,t_4 \}$~\cite{rhim2019classification}. 
We assume that the flat band is at zero energy, and the crossing point is at $\mathbf{k} =0$ without loss of generality.
With $P_0$, the maximum quantum distance $d_\mathrm{max}$ is evaluated as
\begin{eqnarray}
d_\mathrm{max}^2 = \frac{t_4^2}{-t_2^2+4 t_1 t_3 + 2t_4^2}, \label{eq:dmax}
\end{eqnarray}
which is the highest value of $d^2_\mathrm{HS}(\mathbf{k},\mathbf{k}^\prime)$ around the band-crossing point~\cite{rhim2020quantum}. 
Note that the Hamiltonian (\ref{eq:Ham}) is of the form $k^2 h(\phi)$, where $k^2=k_x^2+k_y^2$, and $h$ is a $2\times 2$ matrix and a function of the polar angle with respect to the band-crossing point $\phi$.
As a result, the quantum distance is independent of $k$ because the eigenvectors of (\ref{eq:Ham}) is determined by the $k$-independent part $h(\phi)$.
However, when we calculate $d_\mathrm{max}$ in lattice models, we should take the limit $\mathbf{k},\mathbf{k}^\prime \rightarrow \mathbf{k}_c$, where $\mathbf{k}_c$ is the band-crossing point.
Since the Landau level spreading of the singular flat band, which is arising from the inter-band coupling, is a monotonic function of $d_\mathrm{max}$, $d_\mathrm{max}$ is considered representing the strength of the inter-band coupling between the flat and parabolic bands~\cite{rhim2020quantum}.
One can visualize $d_\mathrm{max}$ by the pseudospin texture $\v s(\v k) = \sum_\alpha \langle \psi_{\v k} |\sigma_\alpha  |  \psi_{\v k} \rangle \hat{\alpha}$, where $\hat{\alpha}$ is a unit vector along the $\alpha$-axis, because the maximum canting angle between the pseudospins $\Delta\theta_\mathrm{max}$ is related to $d_\mathrm{max}$ via an identity $\Delta\theta_\mathrm{max} = 2\sin^{-1}d_\mathrm{max}$~\cite{rhim2020quantum}.

One can transform $P_0$ into another set of gauge-invariant band parameters $P_1=\{ m_{xx}, m_{xy}, m_{yy}, \tilde{d}_{\text{max}} \}$, where $m_{ij}$ is a mass tensor for the quadratic band $E_\mathrm{quad} (\bold{k})$ defined by $m^{-1}_{ij}= \partial_{k_i}\partial_{k_j} E_\mathrm{quad} (\bold{k})|_{\bold{k}\rightarrow 0}$~\cite{rhim2020quantum}.
Here, $\tilde{d}_{\text{max}}= \mathrm{sgn}[t_4]d_{\text{max}}$, where $\mathrm{sgn}[t_4]$ is the sign of $t_4$, can be regarded as an extended definition of the maximum quantum distance, introduced to recover the information of the sign of $t_4$ lost during the transformation (\ref{eq:dmax}).
$P_1$ is the most natural choice of band parameters for the quadratic band-crossing flat band model because three mass tensor elements describe the shape of the parabolic band and $d_{\text{max}}$ corresponds to the interband coupling. 
Moreover, both the mass tensor and $d_{\text{max}}$ are gauge-invariant.
This continuum Hamiltonian (\ref{eq:Ham}) describes the low-energy physics of a variety of well-known flat band models, such as kagome and checkerboard lattice models.
While flat bands are mostly trivial from a topological perspective, it was noted that $d_{\text{max}}$ is the characteristic geometric quantity of the flat band model $\mathcal{H}_{\mathrm{fb}}(\bold{k})$, and manifests itself in Landau level structures~\cite{rhim2020quantum}.

In this work, we focus on the flat band systems because they manifest the effect of $d_\mathrm{max}$ most effectively.
The flat band Hamiltonian (\ref{eq:Ham}) is the most minimal quadratic $2\times 2$ model described by only four independent parameters as explained above.
Except three parameters relevant to the mass tensor or the shape of the parabolic band, only one parameter, $d_\mathrm{max}$, is left to represent the inter-band coupling between the flat and quadratic bands as a geometric quantity.
Moreover, since the flat band model (\ref{eq:Ham}) is topologically trivial~\cite{rhim2020quantum}, the flat band model (\ref{eq:Ham}) is an ideal platform to study geometric phenomena excluding topological origins.

\textbf{Results}
We show that the bulk number $d_{\text{max}}$ of the flat band is reflected in the electronic structur of the wavefunction of the interface mode.
The interface is realized by applying a position-dependent potential $U(x)$ to the system, where $U(\infty)=V^\mathcal{R}_G$ and $U(-\infty)=V^\mathcal{L}_G$ [see Fig.~\ref{fig:main}a]. 
This potential divides the system into two bulk regions, denoted by $\mathcal{L}$- and $\mathcal{R}$-regions.
Under this condition, where the translational symmetry along the $y$-direction still exists, the flat band Hamiltonian (\ref{eq:Ham}) transforms to
\begin{eqnarray}
\mathcal{H}_{\mathrm{fb,IF}}(k_y) = \mathcal{H}_{\mathrm{fb}}(-i\partial_x,k_y) + U(x)\sigma_0, \label{eq:Ham_interface}
\end{eqnarray}
where, for convenience, we assume $V^\mathcal{L}_G = 0$ and $V^\mathcal{R}_G=U_0$.
In this study, we assume that $U_0<0$ without loss of generality.
The interface state is localized around the boundary between the $\mathcal{R}$- and $\mathcal{L}$-regions, and the corresponding band is located between the parabolic and flat bands.
Namely, $k_y^2/2 m_{yy}+U_0 > \mathcal{E}_\mathrm{IF}(k_y) > U_0$, where $\mathcal{E}_\mathrm{IF}(k_y)$ is the energy dispersion of the interface mode, and $U_0$ is the flat band's energy in the $\mathcal{R}$-region far away from the interface.
Assuming that $U(x)$ is slowly varying ($U^\prime(x) \approx 0$) for $|x|\gg 1$, the energy dispersion of the interface mode near the band-crossing point is obtained as
\begin{eqnarray}
    \mathcal{E}_{\text{IF}}(k_y,U_0) =&& \frac{d_{\text{max}}^2}{2 m_{yy}} k_y^2 + U_0,\label{eq:E_IF}
\end{eqnarray}
and the corresponding eigenfunction is given by 
    \begin{eqnarray}
        \psi_{\mathrm{IF},k_y}(x) \propto \text{exp}\left(-i\eta k_y x\right)&&\text{exp}\left[ s(x)\int^x \lambda(x') dx'\right],\label{eq:interface_mode}
    \end{eqnarray}
where $\eta_{k_y} = -m_{xy}/m_{yy}$, $s(x)=+1(-1)$ for $x<0(x>0)$, and $\lambda(x)= (2/m_{xx}^{-1})^{1/2}[k_y^2/2 m_{yy} +U(x) -\mathcal{E}_{\text{IF}}(k_y)]^{1/2}$[see Supplementary Note 2].
Here, $\lambda(x)$, which is related to the energy difference between the parabolic and interface bands, determines the decay length of the localized mode, whereas $s(x)$ represents the exponential decay of the interface mode away from the interface.
Since the above results are obtained from the behavior of the interface mode far from the interface, where $U^\prime(x)\approx 0$, they are independent of the detailed shape of the interface potential around the interface[see Supplementary Note 2].
We note that there are no interface states when $d_\text{max}=0$(non-singular flat band).

As shown in (\ref{eq:E_IF}), the maximum quantum distance of the bulk system is directly and clearly manifested in the effective mass of the interface mode($m^* = m_{yy}/d^2_\mathrm{max}$).
As $d_\mathrm{max}$ increases from 0 to 1, the interface band emerges out of the flat band and then approaches the parabolic band by lowering its effective mass [see Fig.~\ref{fig:main}c].
This means that one can extract the information of the quantum distance of the bulk by observing the spectrum of the interface state generated between two bulks with different gate voltages.
Another role of $d_\mathrm{max}$ is to determine the localization length of the interface mode via $\lambda(x)$ in (\ref{eq:interface_mode}).
As $d_\mathrm{max}$ increases the localization length also increases because the character of the interface mode resembles the extended feature of the wave function of the bulk parabolic band as the interface band gets closer to the bulk parabolic band with increasing $d_\mathrm{max}$.
On the other hand, as $d_\mathrm{max}$ decreases, the localization length decreases too since the interface band approaches the flat band, which is characterized by the compact localized eigenstates~\cite{rhim2019classification}.

\begin{figure}[t]
\includegraphics[width=85mm]{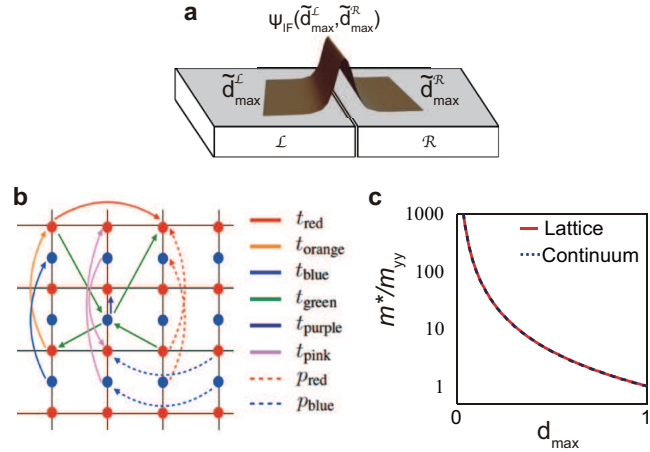} 
\caption{\label{fig4}
\textbf{Interface mode induced by the sign difference of $\tilde{d}_{\text{max}}$.}
\textbf{a}, A junction between two singular flat band systems with different values of $\tilde{d}_{\text{max}}$, which is the extended maximum quantum distance. Here, the onsite potential is the same for both sides.
\textbf{b}, The lattice and hopping structure of the square lattice model hosting a flat band. Here, $t$'s and $p$'s are the hopping parameters. In this model, $d_{\mathrm{max}}$ can be varied from 0 to 1 while maintaining the band-flatness by changing the hopping parameters. [See Supplementary Note 4]. 
\textbf{c}, The $d_{\text{max}}$ dependence of the effective mass of the interface mode where the step-like sign change of $d_{\text{max}}$ is applied, satisfying $\tilde{d}_{\text{max}}(x<0)=-d_{\text{max}}$ and $\tilde{d}_{\text{max}}(x>0)=d_{\text{max}}$.
The generic continuum analysis is represented by the dashed line, which is compared with the results obtained from square lattice model illustrated in \textbf{b}.
Here, $(m^{-1}_{xx},m^{-1}_{xy},m^{-1}_{yy})=(2,0,2)$ is used for the mass tensor.}\label{fig:dmax_diff}
\end{figure}

We also consider another type of interface, where we have two different flat band models on the $\mathcal{L}$- and $\mathcal{R}$-regions, while $d_\mathrm{max}$ is the same on both sides and the applied potential is uniform ($U(x)=0$).
In particular, we obtain analytic results when the two models are the same, except for the sign of $t_4$, which are described by two different parameter sets $\{m_{xx},m_{xy},m_{yy},-d_{\text{max}}\}$ and $\{m_{xx},m_{xy},m_{yy},d_{\text{max}}\}$, respectively [see Fig.~\ref{fig:dmax_diff}a].
Similar to the above, we find that the band spectrum of the interface mode induced by this type of junction is given by $\mathcal{E}_\mathrm{IF}(k_y,0)$.
While we have obtained analytic results only when two regions have the same mass tensor, we note that the interface mode appears even when the mass tensor is different between two regions if $d_{\text{max}}$ is the same for both domains and nonzero.
%

\begin{figure}[b]
\includegraphics[width=83mm]{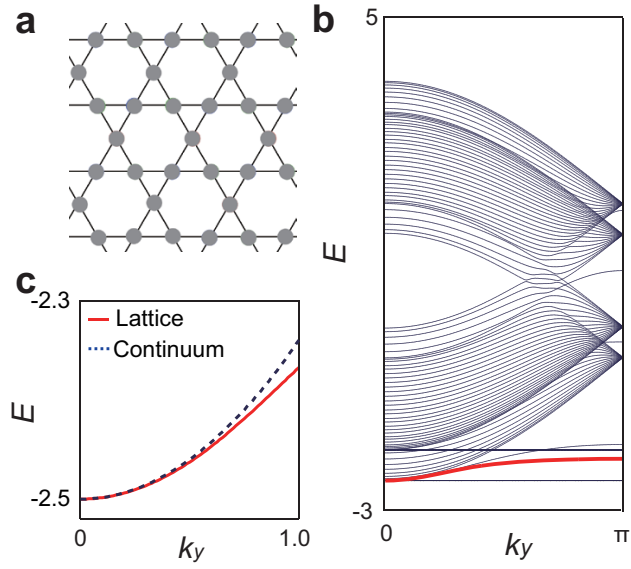} 
\caption{\label{fig2}
\textbf{Interface modes of kagome lattice model.}
\textbf{a}, Kagome lattice.
\textbf{b}, The band structure of the kagome lattice model, where the step-like interface potential $U(x)$ satisfying $U(x<0)=-2$ and $U(x>0)=-2.5$ is applied. The red curve represents the interface states. The other edge bands are from open boundaries.
\textbf{c}, The interface band in \textbf{b} is highlighted and compared with the continuum result (\ref{eq:E_IF}) with $d_\mathrm{max}=1$ and $m^{-1}_{xx}=m^{-1}_{yy}=1/2$, $m^{-1}_{xy}=0$ (dashed line).
}\label{fig:kagome}
\end{figure}

\begin{figure*}[t]
\centering
\includegraphics[width=100mm]{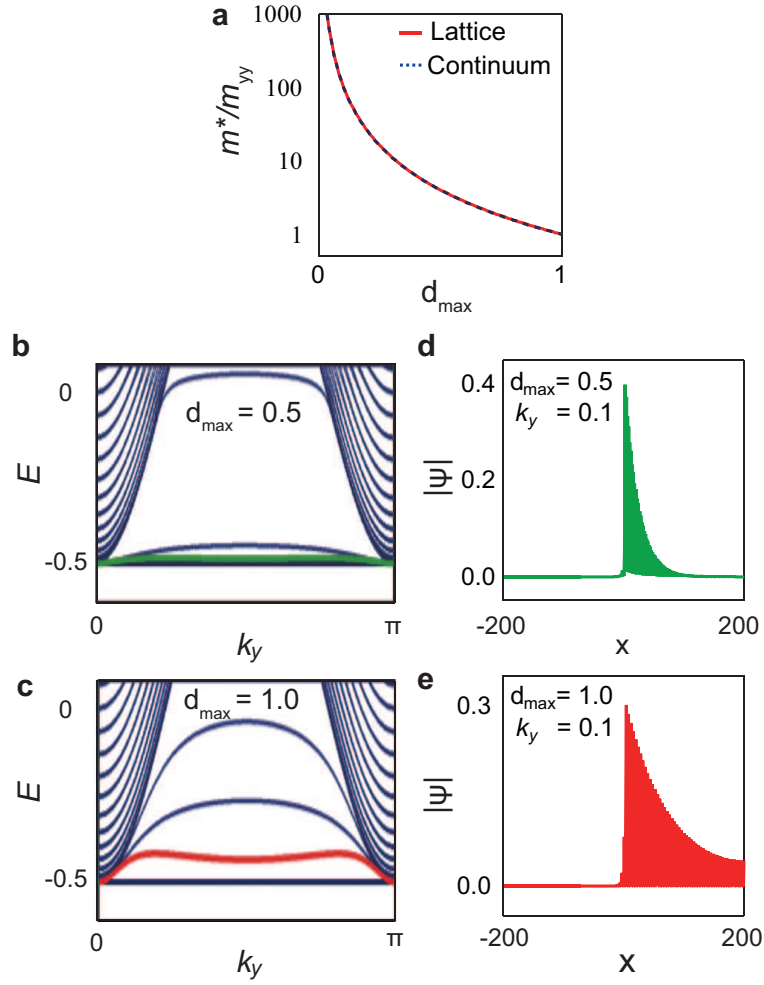} 
\caption{\label{fig3}
\textbf{Band structure and interface modes of the square lattice flat band model.}
\textbf{a}, The effective mass of the interface mode calculated from the lattice and continuum models are plotted as a function of $d_{\mathrm{max}}$. 
\textbf{b} and \textbf{c}, Band structures of the square lattice models at $d_{\mathrm{max}}=0.5$(\textbf{c}) and 1(\textbf{d}), where an interface is introduced by applying an onsite potential $U_0=-0.5$ for $x>0$. Interface bands are represented by green and red colours. The other edge bands are from open boundaries, which are not the focus of this study.
In \textbf{d} and \textbf{e}, the wavefunctions of interface modes at $k_y=0.1$ are plotted for $d_{\mathrm{max}}=0.5$ and 1 respectively.
In all figures, the mass tensor for the parabolic band is fixed as $m^{-1}_{xx}=m^{-1}_{yy}=2$, and $m^{-1}_{xy}=0$.
}\label{fig:square}
\end{figure*}

We apply our general results of the continuum model to various lattice models, such as the kagome, Lieb, and square lattice models.
We consider the conventional tight-binding model on the kagome lattice, where only the nearest-neighbor hopping processes are allowed [see Fig.~\ref{fig:kagome}a]. 
On the other hand, the flat band model on the Lieb lattice is a modified version of the conventional one, where hopping processes are anisotropic, and the onsite potential is non-uniform [see Supplementary Fig.~1a].
Finally, in the case of the square lattice, various artificial long-range hopping processes are included in addition to the nearest neighbor ones of the conventional model as illustrated in Fig.~\ref{fig:square}a.
The hopping parameters are given by $t_{\text{red}}=-t_1/2$, $t_{\text{orange}}=-t_3/2-t_4^2/8t_1$, $t_{\text{green}}=it_4/4$, $t_{\text{blue}}=-t_4^2/8t_1$, $t_{\text{purple}}=t_4(4t_1t_3+t_4^2)^{1/2}/4t_1$, $t_{\text{pink}}=-t_4(4t_1t_3+t_4^2)^{1/2}/8t_1$.
The explicit forms of the tight-binding Hamiltonians for these models are provided in Supplementary Note 4.
They all host a flat band with a quadratic band-crossing at $\mathbf{k}=0$.
While $d_\mathrm{max}$ is fixed to 1 in the kagome and Lieb models, we can control $d_\mathrm{max}$ continuously while maintaining the band flatness and the mass tensor of the quadratic band in the square lattice model.
To investigate the interface modes, we create an interface by applying a step-like potential $U(x)=U_\mathrm{step}(x)$ along $x$-axis such that $U_\mathrm{step}(x)=0$ and $U_\mathrm{step}(x)=U_0$ when $x<0$ and $x>0$, respectively.
We apply an open boundary condition at the left and right ends of the system.

As shown in Fig.~\ref{fig:kagome}b, \ref{fig:square}c and 4d, there are non-bulk bands between the parabolic and flat bands.
The red and green ones indicate the interface modes, while the others represent edge modes localized at the open boundaries of the systems.
The continuum formula (\ref{eq:E_IF}) fits all the interface bands of the lattice models accurately near the band-crossing point.
In the kagome and Lieb lattices, where $d_\mathrm{max}$ is fixed to 1, we confirm that the continuum formula (\ref{eq:E_IF}) works well as shown in the $k_y$-dependence of the interface bands in Fig.~\ref{fig:kagome}c and Supplementary Fig.~1c.
Note that, in the case of the kagome lattice, we apply the interface potential of the form $(U(-\infty),U(\infty))=(-2,-2+U_0)$ to the continuum Hamiltonian (\ref{eq:Ham_interface}) instead of $(U(-\infty),U(\infty)=(0,U_0)$ because the flat band in the kagome lattice model is located at $E=-2$.
In the case of the square lattice model, we additionally plot the $d_\mathrm{max}$-dependence of the effective mass of the interface band by controlling the hopping parameters and then check that the result (\ref{eq:E_IF}) holds [see Fig.~\ref{fig:square}b].
Moreover, the $d_\mathrm{max}$-dependence of the localization length of the interface modes is also illustrated in Fig~\ref{fig:square}e and f, which shows that the wavefunctions becomes extended as $d_\mathrm{max}$ increases.

We examine a junction between two areas with different values of $\tilde{d}_{\text{max}}$ in the square lattice model too [see Fig.~\ref{fig:dmax_diff}a].
We consider a step-like sign change of $\tilde{d}_{\text{max}}$ such that $\tilde{d}_{\text{max}}(x)= -d_{\text{max}}$ and $\tilde{d}_{\text{max}}(x)= d_{\text{max}}$ when $x<0$ and $x>0$, respectively.
In this system, an interface mode also appears as plotted in Supplementary Fig.~7b. 
As shown in Fig.~2b, the effective mass of the interface band(red solid curve) is well described by $m^* = m_{yy}/d^2_\mathrm{max}$ obtained from the continuum analysis.
Although we have considered only the case where the sign of $\tilde{d}_{\text{max}}$ is different between $\mathcal{L}$- and $\mathcal{R}$-regions, we also show that the interface mode appears quite generally when the values of $\tilde{d}_{\text{max}}$ is different between two areas [see Supplementary Note 7 and Supplementary Fig.~7].

We investigate how the bulk-interface correspondence is modified for the nearly flat bands from the perspective of realistic systems.
The interface is introduced by applying the same potential $U(x)$ as we did for the perfectly flat band case.
First, we derive an analytic formula for the dispersion relation of the interface mode and show how it is related to $d_\mathrm{max}$ by considering a continuum Hamiltonian.
The bulk Hamiltonian for the nearly flat band is given by $\mathcal{H}_{\text{n-fb}}(\mathbf{k}) = \mathcal{H}_\mathrm{fb}(\mathbf{k}) + a k_x^2 + b k_y^2 + \Delta\sigma_z$, where $a k_x^2 + b k_y^2$ is a perturbation deforming the flat band and $\Delta\sigma_z$ is the mass term.
When $\Delta=0$ and $a$ or $b$ are nonzero, the band-crossing point survives, and the eigenstates around it give the same $d_\mathrm{max}$, although the flat band is slightly warped.
On the other hand, if $\Delta> 0$($\Delta< 0$), the band-touching point is gapped out when $0 \leq d_\mathrm{max}<1$($0 < d_\mathrm{max}<1$).
When $d_{\text{max}}>0$, that is, when the flat band is singular, this gap-opening process accompanies the warping of the flat band, realizing a nearly flat band.
Although the band-crossing point is destroyed for the nonzero $\Delta$, the effect of $d_\mathrm{max}$ before opening the gap is still encoded in the interface mode for the nearly flat band cases.
When $\Delta> 0$, the energy spectrum of the interface mode around the band-crossing point induced by $U(x)$ is obtained as follows:
\begin{eqnarray}
    \tilde{\mathcal{E}}_{\text{IF}}(k_y,U_0,a,b,\Delta) =  \mathcal{E}_{\text{IF}}(k_y,U_0) + \chi_{ab} k_y^2 - \Delta,\label{eq:nearly_flat_interface_band}
\end{eqnarray}
where $\chi_{ab} = b- a(m_{xx}m_{yy}-\mathrm{det}[\mathbf{m}]d_\mathrm{max}^2)/m_{yy}^2$, as derived in Supplementary Note 3.
Note that the mass tensor used here is from the unperturbed Hamiltonian, $\mathcal{H}_\mathrm{fb}(\mathbf{k})$.
One can extract $d_\mathrm{max}$ from this result assuming that the mass tensor and perturbation parameters are already known from the bulk band structure.
For another type of the interface formed between two regions with opposite signs of $d_{\text{max}}$ but with the same external potential, the band structure of the interface state is found to be equal to $\tilde{\mathcal{E}}_{\text{IF}}^{}(k_y,0,a,b,\Delta)$ similar to the perfectly flat band case.

We confirm that the geometric formula for the interface band (\ref{eq:nearly_flat_interface_band}) works well by examining the square lattice model under the step-like potential $U(x)$.
A nearly flat band is realized in this model by adding several perturbative hopping processes and modulating onsite potentials.
The dashed arrows in Fig.~\ref{fig:square}a represent the perturbative hopping processes: $\{p_{\text{red}},p_{\text{blue}}\}=\{ -a/4,-b/4 \}$.
The explicit form of the corresponding Hamiltonian is provided in Supplementary Note 4.
We show how the continuum formula (\ref{eq:nearly_flat_interface_band}) matches well with the lattice model as a function of crystal momentum and $d_\mathrm{max}$ for the gapless and gapped cases of the square lattice model with the nearly flat band in Fig.~\ref{fig:nearly_flat}(a,b) and Fig.~\ref{fig:nearly_flat}(c,d), respectively [see Supplementary Fig.~2 for the d$_\mathrm{max}$-dependence].
Therefore, one can extract the bulk geometric quantity $d_\mathrm{max}$ of the unperturbed flat band model from the band spectrum of the interface mode of the nearly flat band system near the band-crossing point via the bulk-boundary correspondence described by (\ref{eq:nearly_flat_interface_band}).

\textbf{Discussion}
We have shown that geometric quantities other than topological notions also can be used for the bulk-interface correspondence.
Although our bulk-interface correspondence from the quantum distance applies to flat band systems only, it possesses a certain level of robustness in that it can predict the existence and effective mass of the interface band around the band-crossing point regardless of the explicit shape of the interface potential [see Supplementary Fig.~3, Fig.~4, and Fig.~5 in Supplementary Note 5].
As a result, one can extract the geometric quantity $d_\mathrm{max}$ by observing the effective mass of the interface band even for the nearly flat band cases.
It is noteworthy to mention that the flatness condition is crucial for the existence of the $d_\mathrm{max}$-originating interface mode.
We have shown in Supplementary Note 6 that the interface mode is destroyed when the bandwidth of the nearly flat band becomes comparable to the potential difference $U_0$ [see Supplementary Fig.~6].
The generalization of our bulk-interface correspondence to the band-crossings between generic dispersive bands would be a future research direction, establishing a geometric mechanism for the generation of the interface modes beyond the topological origin.

While the interface is introduced between two regions with different potentials, this kind of experimental setup can be realized by designing a dual-gate device, where two independent voltages are applied to two divided areas of the sample~\cite{choi2011transformable,jeon2015steep}.
The band spectrum of this 1D interface system can then be analyzed from the quasi-particle interference (QPI) pattern obtained from the scanning tunneling microscopy (STM)~\cite{drozdov2014one,zhu2020tunable}.
The anisotropic scattering potential due to the interface formed along the interface lets the QPI propagate mainly into the bulk perpendicular to the interface. 
However, the interface mode might exhibit the charge modulations along the interface so that one can distinguish the QPI of the interface mode from that of the bulk.
If multiple dual-gates with interfaces aligned with each other can be fabricated, the same number of uniformly oriented 1D interface systems can be generated.
The angle-resolved photoemission spectroscopy (ARPES) technique can also be used to observe the band structure of the interface states in such 1D systems~\cite{ahn2005coexistence,senkovskiy2021tunneling,karakachian2020one}. 
Recently, various kagome materials hosting nearly flat bands such as Fe$_3$Sn$_2$, FeSn, and CoSn have been synthesized~\cite{lin2018flatbands,kang2020dirac,kang2020topological}, and they are candidate systems to demonstrate the bulk-interface correspondence from the quantum distance experimentally.
We expect that one of the possible applications utilizing the bulk-interface correspondence is the detection of the ferroelectric domains in memory devices~\cite{scott2000,ma2018controllable,jiang2018temporary}.
Using the interface mode of two-dimensional flat band systems, we propose a quantum device that can detect the domain wall formed with the out-of-plane polarization direction [see Supplementary Fig.~8 in Supplementary note 8].
\\
\\
\textbf{References}

\begin{thebibliography}{56}%
\makeatletter
\providecommand \@ifxundefined [1]{%
 \@ifx{#1\undefined}
}%
\providecommand \@ifnum [1]{%
 \ifnum #1\expandafter \@firstoftwo
 \else \expandafter \@secondoftwo
 \fi
}%
\providecommand \@ifx [1]{%
 \ifx #1\expandafter \@firstoftwo
 \else \expandafter \@secondoftwo
 \fi
}%
\providecommand \natexlab [1]{#1}%
\providecommand \enquote  [1]{``#1''}%
\providecommand \bibnamefont  [1]{#1}%
\providecommand \bibfnamefont [1]{#1}%
\providecommand \citenamefont [1]{#1}%
\providecommand \href@noop [0]{\@secondoftwo}%
\providecommand \href [0]{\begingroup \@sanitize@url \@href}%
\providecommand \@href[1]{\@@startlink{#1}\@@href}%
\providecommand \@@href[1]{\endgroup#1\@@endlink}%
\providecommand \@sanitize@url [0]{\catcode `\\12\catcode `\$12\catcode
  `\&12\catcode `\#12\catcode `\^12\catcode `\_12\catcode `\%12\relax}%
\providecommand \@@startlink[1]{}%
\providecommand \@@endlink[0]{}%
\providecommand \url  [0]{\begingroup\@sanitize@url \@url }%
\providecommand \@url [1]{\endgroup\@href {#1}{\urlprefix }}%
\providecommand \urlprefix  [0]{URL }%
\providecommand \Eprint [0]{\href }%
\providecommand \doibase [0]{http://dx.doi.org/}%
\providecommand \selectlanguage [0]{\@gobble}%
\providecommand \bibinfo  [0]{\@secondoftwo}%
\providecommand \bibfield  [0]{\@secondoftwo}%
\providecommand \translation [1]{[#1]}%
\providecommand \BibitemOpen [0]{}%
\providecommand \bibitemStop [0]{}%
\providecommand \bibitemNoStop [0]{.\EOS\space}%
\providecommand \EOS [0]{\spacefactor3000\relax}%
\providecommand \BibitemShut  [1]{\csname bibitem#1\endcsname}%
\let\auto@bib@innerbib\@empty
\bibitem [{\citenamefont {Kane}\ and\ \citenamefont
  {Mele}(2005{\natexlab{a}})}]{kane2005quantum}%
  \BibitemOpen
  \bibfield  {author} {\bibinfo {author} {\bibfnamefont {Charles~L}\
  \bibnamefont {Kane}}\ and\ \bibinfo {author} {\bibfnamefont {Eugene~J}\
  \bibnamefont {Mele}},\ }\bibfield  {title} {\enquote {\bibinfo {title}
  {Quantum spin hall effect in graphene},}\ }\href@noop {} {\bibfield
  {journal} {\bibinfo  {journal} {Physical review letters}\ }\textbf {\bibinfo
  {volume} {95}},\ \bibinfo {pages} {226801} (\bibinfo {year}
  {2005}{\natexlab{a}})}\BibitemShut {NoStop}%
\bibitem [{\citenamefont {Kane}\ and\ \citenamefont
  {Mele}(2005{\natexlab{b}})}]{kane2005z}%
  \BibitemOpen
  \bibfield  {author} {\bibinfo {author} {\bibfnamefont {Charles~L}\
  \bibnamefont {Kane}}\ and\ \bibinfo {author} {\bibfnamefont {Eugene~J}\
  \bibnamefont {Mele}},\ }\bibfield  {title} {\enquote {\bibinfo {title} {Z 2
  topological order and the quantum spin hall effect},}\ }\href@noop {}
  {\bibfield  {journal} {\bibinfo  {journal} {Physical review letters}\
  }\textbf {\bibinfo {volume} {95}},\ \bibinfo {pages} {146802} (\bibinfo
  {year} {2005}{\natexlab{b}})}\BibitemShut {NoStop}%
\bibitem [{\citenamefont {Bernevig}\ and\ \citenamefont
  {Zhang}(2006)}]{bernevig2006quantum}%
  \BibitemOpen
  \bibfield  {author} {\bibinfo {author} {\bibfnamefont {B~Andrei}\
  \bibnamefont {Bernevig}}\ and\ \bibinfo {author} {\bibfnamefont {Shou-Cheng}\
  \bibnamefont {Zhang}},\ }\bibfield  {title} {\enquote {\bibinfo {title}
  {Quantum spin hall effect},}\ }\href@noop {} {\bibfield  {journal} {\bibinfo
  {journal} {Physical review letters}\ }\textbf {\bibinfo {volume} {96}},\
  \bibinfo {pages} {106802} (\bibinfo {year} {2006})}\BibitemShut {NoStop}%
\bibitem [{\citenamefont {Bernevig}\ \emph {et~al.}(2006)\citenamefont
  {Bernevig}, \citenamefont {Hughes},\ and\ \citenamefont
  {Zhang}}]{bernevig2006quantum_science}%
  \BibitemOpen
  \bibfield  {author} {\bibinfo {author} {\bibfnamefont {B~Andrei}\
  \bibnamefont {Bernevig}}, \bibinfo {author} {\bibfnamefont {Taylor~L}\
  \bibnamefont {Hughes}}, \ and\ \bibinfo {author} {\bibfnamefont {Shou-Cheng}\
  \bibnamefont {Zhang}},\ }\bibfield  {title} {\enquote {\bibinfo {title}
  {Quantum spin hall effect and topological phase transition in hgte quantum
  wells},}\ }\href@noop {} {\bibfield  {journal} {\bibinfo  {journal}
  {Science}\ }\textbf {\bibinfo {volume} {314}},\ \bibinfo {pages} {1757--1761}
  (\bibinfo {year} {2006})}\BibitemShut {NoStop}%
\bibitem [{\citenamefont {Fu}\ \emph {et~al.}(2007)\citenamefont {Fu},
  \citenamefont {Kane},\ and\ \citenamefont {Mele}}]{fu2007topological}%
  \BibitemOpen
  \bibfield  {author} {\bibinfo {author} {\bibfnamefont {Liang}\ \bibnamefont
  {Fu}}, \bibinfo {author} {\bibfnamefont {Charles~L}\ \bibnamefont {Kane}}, \
  and\ \bibinfo {author} {\bibfnamefont {Eugene~J}\ \bibnamefont {Mele}},\
  }\bibfield  {title} {\enquote {\bibinfo {title} {Topological insulators in
  three dimensions},}\ }\href@noop {} {\bibfield  {journal} {\bibinfo
  {journal} {Physical review letters}\ }\textbf {\bibinfo {volume} {98}},\
  \bibinfo {pages} {106803} (\bibinfo {year} {2007})}\BibitemShut {NoStop}%
\bibitem [{\citenamefont {Thouless}\ \emph {et~al.}(1982)\citenamefont
  {Thouless}, \citenamefont {Kohmoto}, \citenamefont {Nightingale},\ and\
  \citenamefont {den Nijs}}]{thouless1982quantized}%
  \BibitemOpen
  \bibfield  {author} {\bibinfo {author} {\bibfnamefont {David~J}\ \bibnamefont
  {Thouless}}, \bibinfo {author} {\bibfnamefont {Mahito}\ \bibnamefont
  {Kohmoto}}, \bibinfo {author} {\bibfnamefont {M~Peter}\ \bibnamefont
  {Nightingale}}, \ and\ \bibinfo {author} {\bibfnamefont {Marcel}\
  \bibnamefont {den Nijs}},\ }\bibfield  {title} {\enquote {\bibinfo {title}
  {Quantized hall conductance in a two-dimensional periodic potential},}\
  }\href@noop {} {\bibfield  {journal} {\bibinfo  {journal} {Physical review
  letters}\ }\textbf {\bibinfo {volume} {49}},\ \bibinfo {pages} {405}
  (\bibinfo {year} {1982})}\BibitemShut {NoStop}%
\bibitem [{\citenamefont {Moore}\ and\ \citenamefont
  {Balents}(2007)}]{moore2007topological}%
  \BibitemOpen
  \bibfield  {author} {\bibinfo {author} {\bibfnamefont {Joel~E}\ \bibnamefont
  {Moore}}\ and\ \bibinfo {author} {\bibfnamefont {Leon}\ \bibnamefont
  {Balents}},\ }\bibfield  {title} {\enquote {\bibinfo {title} {Topological
  invariants of time-reversal-invariant band structures},}\ }\href@noop {}
  {\bibfield  {journal} {\bibinfo  {journal} {Physical Review B}\ }\textbf
  {\bibinfo {volume} {75}},\ \bibinfo {pages} {121306} (\bibinfo {year}
  {2007})}\BibitemShut {NoStop}%
\bibitem [{\citenamefont {Schnyder}\ \emph {et~al.}(2008)\citenamefont
  {Schnyder}, \citenamefont {Ryu}, \citenamefont {Furusaki},\ and\
  \citenamefont {Ludwig}}]{schnyder2008classification}%
  \BibitemOpen
  \bibfield  {author} {\bibinfo {author} {\bibfnamefont {Andreas~P}\
  \bibnamefont {Schnyder}}, \bibinfo {author} {\bibfnamefont {Shinsei}\
  \bibnamefont {Ryu}}, \bibinfo {author} {\bibfnamefont {Akira}\ \bibnamefont
  {Furusaki}}, \ and\ \bibinfo {author} {\bibfnamefont {Andreas~WW}\
  \bibnamefont {Ludwig}},\ }\bibfield  {title} {\enquote {\bibinfo {title}
  {Classification of topological insulators and superconductors in three
  spatial dimensions},}\ }\href@noop {} {\bibfield  {journal} {\bibinfo
  {journal} {Physical Review B}\ }\textbf {\bibinfo {volume} {78}},\ \bibinfo
  {pages} {195125} (\bibinfo {year} {2008})}\BibitemShut {NoStop}%
\bibitem [{\citenamefont {Ando}\ and\ \citenamefont
  {Fu}(2015)}]{ando2015topological}%
  \BibitemOpen
  \bibfield  {author} {\bibinfo {author} {\bibfnamefont {Yoichi}\ \bibnamefont
  {Ando}}\ and\ \bibinfo {author} {\bibfnamefont {Liang}\ \bibnamefont {Fu}},\
  }\bibfield  {title} {\enquote {\bibinfo {title} {Topological crystalline
  insulators and topological superconductors: From concepts to materials},}\
  }\href@noop {} {\bibfield  {journal} {\bibinfo  {journal} {Annu. Rev.
  Condens. Matter Phys.}\ }\textbf {\bibinfo {volume} {6}},\ \bibinfo {pages}
  {361--381} (\bibinfo {year} {2015})}\BibitemShut {NoStop}%
\bibitem [{\citenamefont {Haldane}(1988)}]{haldane1988model}%
  \BibitemOpen
  \bibfield  {author} {\bibinfo {author} {\bibfnamefont {F~Duncan~M}\
  \bibnamefont {Haldane}},\ }\bibfield  {title} {\enquote {\bibinfo {title}
  {Model for a quantum hall effect without landau levels: Condensed-matter
  realization of the" parity anomaly"},}\ }\href@noop {} {\bibfield  {journal}
  {\bibinfo  {journal} {Physical review letters}\ }\textbf {\bibinfo {volume}
  {61}},\ \bibinfo {pages} {2015} (\bibinfo {year} {1988})}\BibitemShut
  {NoStop}%
\bibitem [{\citenamefont {Teo}\ \emph {et~al.}(2008)\citenamefont {Teo},
  \citenamefont {Fu},\ and\ \citenamefont {Kane}}]{teo2008surface}%
  \BibitemOpen
  \bibfield  {author} {\bibinfo {author} {\bibfnamefont {Jeffrey~CY}\
  \bibnamefont {Teo}}, \bibinfo {author} {\bibfnamefont {Liang}\ \bibnamefont
  {Fu}}, \ and\ \bibinfo {author} {\bibfnamefont {CL}~\bibnamefont {Kane}},\
  }\bibfield  {title} {\enquote {\bibinfo {title} {Surface states and
  topological invariants in three-dimensional topological insulators:
  Application to bi 1- x sb x},}\ }\href@noop {} {\bibfield  {journal}
  {\bibinfo  {journal} {Physical Review B}\ }\textbf {\bibinfo {volume} {78}},\
  \bibinfo {pages} {045426} (\bibinfo {year} {2008})}\BibitemShut {NoStop}%
\bibitem [{\citenamefont {Wan}\ \emph {et~al.}(2011)\citenamefont {Wan},
  \citenamefont {Turner}, \citenamefont {Vishwanath},\ and\ \citenamefont
  {Savrasov}}]{wan2011topological}%
  \BibitemOpen
  \bibfield  {author} {\bibinfo {author} {\bibfnamefont {Xiangang}\
  \bibnamefont {Wan}}, \bibinfo {author} {\bibfnamefont {Ari~M}\ \bibnamefont
  {Turner}}, \bibinfo {author} {\bibfnamefont {Ashvin}\ \bibnamefont
  {Vishwanath}}, \ and\ \bibinfo {author} {\bibfnamefont {Sergey~Y}\
  \bibnamefont {Savrasov}},\ }\bibfield  {title} {\enquote {\bibinfo {title}
  {Topological semimetal and fermi-arc surface states in the electronic
  structure of pyrochlore iridates},}\ }\href@noop {} {\bibfield  {journal}
  {\bibinfo  {journal} {Physical Review B}\ }\textbf {\bibinfo {volume} {83}},\
  \bibinfo {pages} {205101} (\bibinfo {year} {2011})}\BibitemShut {NoStop}%
\bibitem [{\citenamefont {Zak}(1989)}]{zak1989berry}%
  \BibitemOpen
  \bibfield  {author} {\bibinfo {author} {\bibfnamefont {J}~\bibnamefont
  {Zak}},\ }\bibfield  {title} {\enquote {\bibinfo {title} {Berry’s phase for
  energy bands in solids},}\ }\href@noop {} {\bibfield  {journal} {\bibinfo
  {journal} {Physical review letters}\ }\textbf {\bibinfo {volume} {62}},\
  \bibinfo {pages} {2747} (\bibinfo {year} {1989})}\BibitemShut {NoStop}%
\bibitem [{\citenamefont {Atala}\ \emph {et~al.}(2013)\citenamefont {Atala},
  \citenamefont {Aidelsburger}, \citenamefont {Barreiro}, \citenamefont
  {Abanin}, \citenamefont {Kitagawa}, \citenamefont {Demler},\ and\
  \citenamefont {Bloch}}]{atala2013direct}%
  \BibitemOpen
  \bibfield  {author} {\bibinfo {author} {\bibfnamefont {Marcos}\ \bibnamefont
  {Atala}}, \bibinfo {author} {\bibfnamefont {Monika}\ \bibnamefont
  {Aidelsburger}}, \bibinfo {author} {\bibfnamefont {Julio~T}\ \bibnamefont
  {Barreiro}}, \bibinfo {author} {\bibfnamefont {Dmitry}\ \bibnamefont
  {Abanin}}, \bibinfo {author} {\bibfnamefont {Takuya}\ \bibnamefont
  {Kitagawa}}, \bibinfo {author} {\bibfnamefont {Eugene}\ \bibnamefont
  {Demler}}, \ and\ \bibinfo {author} {\bibfnamefont {Immanuel}\ \bibnamefont
  {Bloch}},\ }\bibfield  {title} {\enquote {\bibinfo {title} {Direct
  measurement of the zak phase in topological bloch bands},}\ }\href@noop {}
  {\bibfield  {journal} {\bibinfo  {journal} {Nature Physics}\ }\textbf
  {\bibinfo {volume} {9}},\ \bibinfo {pages} {795--800} (\bibinfo {year}
  {2013})}\BibitemShut {NoStop}%
\bibitem [{\citenamefont {Rhim}\ \emph {et~al.}(2017)\citenamefont {Rhim},
  \citenamefont {Behrends},\ and\ \citenamefont {Bardarson}}]{rhim2017bulk}%
  \BibitemOpen
  \bibfield  {author} {\bibinfo {author} {\bibfnamefont {Jun-Won}\ \bibnamefont
  {Rhim}}, \bibinfo {author} {\bibfnamefont {Jan}\ \bibnamefont {Behrends}}, \
  and\ \bibinfo {author} {\bibfnamefont {Jens~H}\ \bibnamefont {Bardarson}},\
  }\bibfield  {title} {\enquote {\bibinfo {title} {Bulk-boundary correspondence
  from the intercellular zak phase},}\ }\href@noop {} {\bibfield  {journal}
  {\bibinfo  {journal} {Physical Review B}\ }\textbf {\bibinfo {volume} {95}},\
  \bibinfo {pages} {035421} (\bibinfo {year} {2017})}\BibitemShut {NoStop}%
\bibitem [{\citenamefont {Hatsugai}(1993)}]{hatsugai1993chern}%
  \BibitemOpen
  \bibfield  {author} {\bibinfo {author} {\bibfnamefont {Yasuhiro}\
  \bibnamefont {Hatsugai}},\ }\bibfield  {title} {\enquote {\bibinfo {title}
  {Chern number and edge states in the integer quantum hall effect},}\
  }\href@noop {} {\bibfield  {journal} {\bibinfo  {journal} {Physical review
  letters}\ }\textbf {\bibinfo {volume} {71}},\ \bibinfo {pages} {3697}
  (\bibinfo {year} {1993})}\BibitemShut {NoStop}%
\bibitem [{\citenamefont {Kitaev}(2009)}]{kitaev2009periodic}%
  \BibitemOpen
  \bibfield  {author} {\bibinfo {author} {\bibfnamefont {Alexei}\ \bibnamefont
  {Kitaev}},\ }\bibfield  {title} {\enquote {\bibinfo {title} {Periodic table
  for topological insulators and superconductors},}\ }in\ \href@noop {} {\emph
  {\bibinfo {booktitle} {AIP conference proceedings}}},\ Vol.\ \bibinfo
  {volume} {1134}\ (\bibinfo {organization} {American Institute of Physics},\
  \bibinfo {year} {2009})\ pp.\ \bibinfo {pages} {22--30}\BibitemShut {NoStop}%
\bibitem [{\citenamefont {Fukui}\ \emph {et~al.}(2012)\citenamefont {Fukui},
  \citenamefont {Shiozaki}, \citenamefont {Fujiwara},\ and\ \citenamefont
  {Fujimoto}}]{fukui2012bulk}%
  \BibitemOpen
  \bibfield  {author} {\bibinfo {author} {\bibfnamefont {Takahiro}\
  \bibnamefont {Fukui}}, \bibinfo {author} {\bibfnamefont {Ken}\ \bibnamefont
  {Shiozaki}}, \bibinfo {author} {\bibfnamefont {Takanori}\ \bibnamefont
  {Fujiwara}}, \ and\ \bibinfo {author} {\bibfnamefont {Satoshi}\ \bibnamefont
  {Fujimoto}},\ }\bibfield  {title} {\enquote {\bibinfo {title} {Bulk-edge
  correspondence for chern topological phases: A viewpoint from a generalized
  index theorem},}\ }\href@noop {} {\bibfield  {journal} {\bibinfo  {journal}
  {Journal of the Physical Society of Japan}\ }\textbf {\bibinfo {volume}
  {81}},\ \bibinfo {pages} {114602} (\bibinfo {year} {2012})}\BibitemShut
  {NoStop}%
\bibitem [{\citenamefont {Mong}\ and\ \citenamefont
  {Shivamoggi}(2011)}]{mong2011edge}%
  \BibitemOpen
  \bibfield  {author} {\bibinfo {author} {\bibfnamefont {Roger~SK}\
  \bibnamefont {Mong}}\ and\ \bibinfo {author} {\bibfnamefont {Vasudha}\
  \bibnamefont {Shivamoggi}},\ }\bibfield  {title} {\enquote {\bibinfo {title}
  {Edge states and the bulk-boundary correspondence in dirac hamiltonians},}\
  }\href@noop {} {\bibfield  {journal} {\bibinfo  {journal} {Physical Review
  B}\ }\textbf {\bibinfo {volume} {83}},\ \bibinfo {pages} {125109} (\bibinfo
  {year} {2011})}\BibitemShut {NoStop}%
\bibitem [{\citenamefont {Prodan}\ and\ \citenamefont
  {Schulz-Baldes}(2016)}]{prodan2016bulk}%
  \BibitemOpen
  \bibfield  {author} {\bibinfo {author} {\bibfnamefont {Emil}\ \bibnamefont
  {Prodan}}\ and\ \bibinfo {author} {\bibfnamefont {Hermann}\ \bibnamefont
  {Schulz-Baldes}},\ }\bibfield  {title} {\enquote {\bibinfo {title} {Bulk and
  boundary invariants for complex topological insulators},}\ }\href@noop {}
  {\bibfield  {journal} {\bibinfo  {journal} {K}\ } (\bibinfo {year}
  {2016})}\BibitemShut {NoStop}%
\bibitem [{\citenamefont {Rhim}\ \emph {et~al.}(2018)\citenamefont {Rhim},
  \citenamefont {Bardarson},\ and\ \citenamefont {Slager}}]{rhim2018unified}%
  \BibitemOpen
  \bibfield  {author} {\bibinfo {author} {\bibfnamefont {Jun-Won}\ \bibnamefont
  {Rhim}}, \bibinfo {author} {\bibfnamefont {Jens~H}\ \bibnamefont
  {Bardarson}}, \ and\ \bibinfo {author} {\bibfnamefont {Robert-Jan}\
  \bibnamefont {Slager}},\ }\bibfield  {title} {\enquote {\bibinfo {title}
  {Unified bulk-boundary correspondence for band insulators},}\ }\href@noop {}
  {\bibfield  {journal} {\bibinfo  {journal} {Physical Review B}\ }\textbf
  {\bibinfo {volume} {97}},\ \bibinfo {pages} {115143} (\bibinfo {year}
  {2018})}\BibitemShut {NoStop}%
\bibitem [{\citenamefont {Bu{\v{z}}ek}\ and\ \citenamefont
  {Hillery}(1996)}]{buvzek1996quantum}%
  \BibitemOpen
  \bibfield  {author} {\bibinfo {author} {\bibfnamefont {Vladimir}\
  \bibnamefont {Bu{\v{z}}ek}}\ and\ \bibinfo {author} {\bibfnamefont {Mark}\
  \bibnamefont {Hillery}},\ }\bibfield  {title} {\enquote {\bibinfo {title}
  {Quantum copying: Beyond the no-cloning theorem},}\ }\href@noop {} {\bibfield
   {journal} {\bibinfo  {journal} {Physical Review A}\ }\textbf {\bibinfo
  {volume} {54}},\ \bibinfo {pages} {1844} (\bibinfo {year}
  {1996})}\BibitemShut {NoStop}%
\bibitem [{\citenamefont {Dodonov}\ \emph {et~al.}(2000)\citenamefont
  {Dodonov}, \citenamefont {Man'Ko}, \citenamefont {Man'Ko},\ and\
  \citenamefont {W{\"u}nsche}}]{dodonov2000hilbert}%
  \BibitemOpen
  \bibfield  {author} {\bibinfo {author} {\bibfnamefont {VV}~\bibnamefont
  {Dodonov}}, \bibinfo {author} {\bibfnamefont {OV}~\bibnamefont {Man'Ko}},
  \bibinfo {author} {\bibfnamefont {VI}~\bibnamefont {Man'Ko}}, \ and\ \bibinfo
  {author} {\bibfnamefont {A}~\bibnamefont {W{\"u}nsche}},\ }\bibfield  {title}
  {\enquote {\bibinfo {title} {Hilbert-schmidt distance and non-classicality of
  states in quantum optics},}\ }\href@noop {} {\bibfield  {journal} {\bibinfo
  {journal} {Journal of Modern Optics}\ }\textbf {\bibinfo {volume} {47}},\
  \bibinfo {pages} {633--654} (\bibinfo {year} {2000})}\BibitemShut {NoStop}%
\bibitem [{\citenamefont {Wilczek}\ and\ \citenamefont
  {Shapere}(1989)}]{wilczek1989geometric}%
  \BibitemOpen
  \bibfield  {author} {\bibinfo {author} {\bibfnamefont {Frank}\ \bibnamefont
  {Wilczek}}\ and\ \bibinfo {author} {\bibfnamefont {Alfred}\ \bibnamefont
  {Shapere}},\ }\href@noop {} {\emph {\bibinfo {title} {Geometric phases in
  physics}}},\ Vol.~\bibinfo {volume} {5}\ (\bibinfo  {publisher} {World
  Scientific},\ \bibinfo {year} {1989})\BibitemShut {NoStop}%
\bibitem [{\citenamefont {Provost}\ and\ \citenamefont
  {Vallee}(1980)}]{provost1980riemannian}%
  \BibitemOpen
  \bibfield  {author} {\bibinfo {author} {\bibfnamefont {JP}~\bibnamefont
  {Provost}}\ and\ \bibinfo {author} {\bibfnamefont {G}~\bibnamefont
  {Vallee}},\ }\bibfield  {title} {\enquote {\bibinfo {title} {Riemannian
  structure on manifolds of quantum states},}\ }\href@noop {} {\bibfield
  {journal} {\bibinfo  {journal} {Communications in Mathematical Physics}\
  }\textbf {\bibinfo {volume} {76}},\ \bibinfo {pages} {289--301} (\bibinfo
  {year} {1980})}\BibitemShut {NoStop}%
\bibitem [{\citenamefont {Ozawa}\ and\ \citenamefont
  {Mera}(2021)}]{ozawa2021relations}%
  \BibitemOpen
  \bibfield  {author} {\bibinfo {author} {\bibfnamefont {Tomoki}\ \bibnamefont
  {Ozawa}}\ and\ \bibinfo {author} {\bibfnamefont {Bruno}\ \bibnamefont
  {Mera}},\ }\bibfield  {title} {\enquote {\bibinfo {title} {Relations between
  topology and the quantum metric for chern insulators},}\ }\href@noop {}
  {\bibfield  {journal} {\bibinfo  {journal} {Physical Review B}\ }\textbf
  {\bibinfo {volume} {104}},\ \bibinfo {pages} {045103} (\bibinfo {year}
  {2021})}\BibitemShut {NoStop}%
\bibitem [{\citenamefont {Mera}\ \emph {et~al.}(2022)\citenamefont {Mera},
  \citenamefont {Zhang},\ and\ \citenamefont {Goldman}}]{mera2022relating}%
  \BibitemOpen
  \bibfield  {author} {\bibinfo {author} {\bibfnamefont {Bruno}\ \bibnamefont
  {Mera}}, \bibinfo {author} {\bibfnamefont {Anwei}\ \bibnamefont {Zhang}}, \
  and\ \bibinfo {author} {\bibfnamefont {Nathan}\ \bibnamefont {Goldman}},\
  }\bibfield  {title} {\enquote {\bibinfo {title} {Relating the topology of
  dirac hamiltonians to quantum geometry: When the quantum metric dictates
  chern numbers and winding numbers},}\ }\href@noop {} {\bibfield  {journal}
  {\bibinfo  {journal} {SciPost Physics}\ }\textbf {\bibinfo {volume} {12}},\
  \bibinfo {pages} {018} (\bibinfo {year} {2022})}\BibitemShut {NoStop}%
\bibitem [{\citenamefont {Morell}\ \emph {et~al.}(2010)\citenamefont {Morell},
  \citenamefont {Correa}, \citenamefont {Vargas}, \citenamefont {Pacheco},\
  and\ \citenamefont {Barticevic}}]{morell2010flat}%
  \BibitemOpen
  \bibfield  {author} {\bibinfo {author} {\bibfnamefont {E~Su{\'a}rez}\
  \bibnamefont {Morell}}, \bibinfo {author} {\bibfnamefont {JD}~\bibnamefont
  {Correa}}, \bibinfo {author} {\bibfnamefont {P}~\bibnamefont {Vargas}},
  \bibinfo {author} {\bibfnamefont {M}~\bibnamefont {Pacheco}}, \ and\ \bibinfo
  {author} {\bibfnamefont {Z}~\bibnamefont {Barticevic}},\ }\bibfield  {title}
  {\enquote {\bibinfo {title} {Flat bands in slightly twisted bilayer graphene:
  Tight-binding calculations},}\ }\href@noop {} {\bibfield  {journal} {\bibinfo
   {journal} {Physical Review B}\ }\textbf {\bibinfo {volume} {82}},\ \bibinfo
  {pages} {121407} (\bibinfo {year} {2010})}\BibitemShut {NoStop}%
\bibitem [{\citenamefont {Cao}\ \emph {et~al.}(2018)\citenamefont {Cao},
  \citenamefont {Fatemi}, \citenamefont {Fang}, \citenamefont {Watanabe},
  \citenamefont {Taniguchi}, \citenamefont {Kaxiras},\ and\ \citenamefont
  {Jarillo-Herrero}}]{cao2018unconventional}%
  \BibitemOpen
  \bibfield  {author} {\bibinfo {author} {\bibfnamefont {Yuan}\ \bibnamefont
  {Cao}}, \bibinfo {author} {\bibfnamefont {Valla}\ \bibnamefont {Fatemi}},
  \bibinfo {author} {\bibfnamefont {Shiang}\ \bibnamefont {Fang}}, \bibinfo
  {author} {\bibfnamefont {Kenji}\ \bibnamefont {Watanabe}}, \bibinfo {author}
  {\bibfnamefont {Takashi}\ \bibnamefont {Taniguchi}}, \bibinfo {author}
  {\bibfnamefont {Efthimios}\ \bibnamefont {Kaxiras}}, \ and\ \bibinfo {author}
  {\bibfnamefont {Pablo}\ \bibnamefont {Jarillo-Herrero}},\ }\bibfield  {title}
  {\enquote {\bibinfo {title} {Unconventional superconductivity in magic-angle
  graphene superlattices},}\ }\href@noop {} {\bibfield  {journal} {\bibinfo
  {journal} {Nature}\ }\textbf {\bibinfo {volume} {556}},\ \bibinfo {pages}
  {43--50} (\bibinfo {year} {2018})}\BibitemShut {NoStop}%
\bibitem [{\citenamefont {Balents}\ \emph {et~al.}(2020)\citenamefont
  {Balents}, \citenamefont {Dean}, \citenamefont {Efetov},\ and\ \citenamefont
  {Young}}]{balents2020superconductivity}%
  \BibitemOpen
  \bibfield  {author} {\bibinfo {author} {\bibfnamefont {Leon}\ \bibnamefont
  {Balents}}, \bibinfo {author} {\bibfnamefont {Cory~R}\ \bibnamefont {Dean}},
  \bibinfo {author} {\bibfnamefont {Dmitri~K}\ \bibnamefont {Efetov}}, \ and\
  \bibinfo {author} {\bibfnamefont {Andrea~F}\ \bibnamefont {Young}},\
  }\bibfield  {title} {\enquote {\bibinfo {title} {Superconductivity and strong
  correlations in moir{\'e} flat bands},}\ }\href@noop {} {\bibfield  {journal}
  {\bibinfo  {journal} {Nature Physics}\ }\textbf {\bibinfo {volume} {16}},\
  \bibinfo {pages} {725--733} (\bibinfo {year} {2020})}\BibitemShut {NoStop}%
\bibitem [{\citenamefont {Peri}\ \emph {et~al.}(2021)\citenamefont {Peri},
  \citenamefont {Song}, \citenamefont {Bernevig},\ and\ \citenamefont
  {Huber}}]{peri2021fragile}%
  \BibitemOpen
  \bibfield  {author} {\bibinfo {author} {\bibfnamefont {Valerio}\ \bibnamefont
  {Peri}}, \bibinfo {author} {\bibfnamefont {Zhi-Da}\ \bibnamefont {Song}},
  \bibinfo {author} {\bibfnamefont {B~Andrei}\ \bibnamefont {Bernevig}}, \ and\
  \bibinfo {author} {\bibfnamefont {Sebastian~D}\ \bibnamefont {Huber}},\
  }\bibfield  {title} {\enquote {\bibinfo {title} {Fragile topology and
  flat-band superconductivity in the strong-coupling regime},}\ }\href@noop {}
  {\bibfield  {journal} {\bibinfo  {journal} {Physical review letters}\
  }\textbf {\bibinfo {volume} {126}},\ \bibinfo {pages} {027002} (\bibinfo
  {year} {2021})}\BibitemShut {NoStop}%
\bibitem [{\citenamefont {Liu}\ \emph {et~al.}(2021)\citenamefont {Liu},
  \citenamefont {Chiu}, \citenamefont {Lee}, \citenamefont {Farahi},
  \citenamefont {Watanabe}, \citenamefont {Taniguchi}, \citenamefont
  {Vishwanath},\ and\ \citenamefont {Yazdani}}]{liu2021spectroscopy}%
  \BibitemOpen
  \bibfield  {author} {\bibinfo {author} {\bibfnamefont {Xiaomeng}\
  \bibnamefont {Liu}}, \bibinfo {author} {\bibfnamefont {Cheng-Li}\
  \bibnamefont {Chiu}}, \bibinfo {author} {\bibfnamefont {Jong~Yeon}\
  \bibnamefont {Lee}}, \bibinfo {author} {\bibfnamefont {Gelareh}\ \bibnamefont
  {Farahi}}, \bibinfo {author} {\bibfnamefont {Kenji}\ \bibnamefont
  {Watanabe}}, \bibinfo {author} {\bibfnamefont {Takashi}\ \bibnamefont
  {Taniguchi}}, \bibinfo {author} {\bibfnamefont {Ashvin}\ \bibnamefont
  {Vishwanath}}, \ and\ \bibinfo {author} {\bibfnamefont {Ali}\ \bibnamefont
  {Yazdani}},\ }\bibfield  {title} {\enquote {\bibinfo {title} {Spectroscopy of
  a tunable moir{\'e} system with a correlated and topological flat band},}\
  }\href@noop {} {\bibfield  {journal} {\bibinfo  {journal} {Nature
  communications}\ }\textbf {\bibinfo {volume} {12}},\ \bibinfo {pages} {1--7}
  (\bibinfo {year} {2021})}\BibitemShut {NoStop}%
\bibitem [{\citenamefont {Sharpe}\ \emph {et~al.}(2019)\citenamefont {Sharpe},
  \citenamefont {Fox}, \citenamefont {Barnard}, \citenamefont {Finney},
  \citenamefont {Watanabe}, \citenamefont {Taniguchi}, \citenamefont
  {Kastner},\ and\ \citenamefont {Goldhaber-Gordon}}]{sharpe2019emergent}%
  \BibitemOpen
  \bibfield  {author} {\bibinfo {author} {\bibfnamefont {Aaron~L}\ \bibnamefont
  {Sharpe}}, \bibinfo {author} {\bibfnamefont {Eli~J}\ \bibnamefont {Fox}},
  \bibinfo {author} {\bibfnamefont {Arthur~W}\ \bibnamefont {Barnard}},
  \bibinfo {author} {\bibfnamefont {Joe}\ \bibnamefont {Finney}}, \bibinfo
  {author} {\bibfnamefont {Kenji}\ \bibnamefont {Watanabe}}, \bibinfo {author}
  {\bibfnamefont {Takashi}\ \bibnamefont {Taniguchi}}, \bibinfo {author}
  {\bibfnamefont {MA}~\bibnamefont {Kastner}}, \ and\ \bibinfo {author}
  {\bibfnamefont {David}\ \bibnamefont {Goldhaber-Gordon}},\ }\bibfield
  {title} {\enquote {\bibinfo {title} {Emergent ferromagnetism near
  three-quarters filling in twisted bilayer graphene},}\ }\href@noop {}
  {\bibfield  {journal} {\bibinfo  {journal} {Science}\ }\textbf {\bibinfo
  {volume} {365}},\ \bibinfo {pages} {605--608} (\bibinfo {year}
  {2019})}\BibitemShut {NoStop}%
\bibitem [{\citenamefont {Wu}\ \emph {et~al.}(2021)\citenamefont {Wu},
  \citenamefont {Zhang}, \citenamefont {Watanabe}, \citenamefont {Taniguchi},\
  and\ \citenamefont {Andrei}}]{wu2021chern}%
  \BibitemOpen
  \bibfield  {author} {\bibinfo {author} {\bibfnamefont {Shuang}\ \bibnamefont
  {Wu}}, \bibinfo {author} {\bibfnamefont {Zhenyuan}\ \bibnamefont {Zhang}},
  \bibinfo {author} {\bibfnamefont {K}~\bibnamefont {Watanabe}}, \bibinfo
  {author} {\bibfnamefont {T}~\bibnamefont {Taniguchi}}, \ and\ \bibinfo
  {author} {\bibfnamefont {Eva~Y}\ \bibnamefont {Andrei}},\ }\bibfield  {title}
  {\enquote {\bibinfo {title} {Chern insulators, van hove singularities and
  topological flat bands in magic-angle twisted bilayer graphene},}\
  }\href@noop {} {\bibfield  {journal} {\bibinfo  {journal} {Nature materials}\
  }\textbf {\bibinfo {volume} {20}},\ \bibinfo {pages} {488--494} (\bibinfo
  {year} {2021})}\BibitemShut {NoStop}%
\bibitem [{\citenamefont {Saito}\ \emph {et~al.}(2021)\citenamefont {Saito},
  \citenamefont {Ge}, \citenamefont {Rademaker}, \citenamefont {Watanabe},
  \citenamefont {Taniguchi}, \citenamefont {Abanin},\ and\ \citenamefont
  {Young}}]{saito2021hofstadter}%
  \BibitemOpen
  \bibfield  {author} {\bibinfo {author} {\bibfnamefont {Yu}~\bibnamefont
  {Saito}}, \bibinfo {author} {\bibfnamefont {Jingyuan}\ \bibnamefont {Ge}},
  \bibinfo {author} {\bibfnamefont {Louk}\ \bibnamefont {Rademaker}}, \bibinfo
  {author} {\bibfnamefont {Kenji}\ \bibnamefont {Watanabe}}, \bibinfo {author}
  {\bibfnamefont {Takashi}\ \bibnamefont {Taniguchi}}, \bibinfo {author}
  {\bibfnamefont {Dmitry~A}\ \bibnamefont {Abanin}}, \ and\ \bibinfo {author}
  {\bibfnamefont {Andrea~F}\ \bibnamefont {Young}},\ }\bibfield  {title}
  {\enquote {\bibinfo {title} {Hofstadter subband ferromagnetism and
  symmetry-broken chern insulators in twisted bilayer graphene},}\ }\href@noop
  {} {\bibfield  {journal} {\bibinfo  {journal} {Nature Physics}\ }\textbf
  {\bibinfo {volume} {17}},\ \bibinfo {pages} {478--481} (\bibinfo {year}
  {2021})}\BibitemShut {NoStop}%
\bibitem [{\citenamefont {Hase}\ \emph {et~al.}(2018)\citenamefont {Hase},
  \citenamefont {Yanagisawa}, \citenamefont {Aiura},\ and\ \citenamefont
  {Kawashima}}]{hase2018possibility}%
  \BibitemOpen
  \bibfield  {author} {\bibinfo {author} {\bibfnamefont {I}~\bibnamefont
  {Hase}}, \bibinfo {author} {\bibfnamefont {T}~\bibnamefont {Yanagisawa}},
  \bibinfo {author} {\bibfnamefont {Y}~\bibnamefont {Aiura}}, \ and\ \bibinfo
  {author} {\bibfnamefont {K}~\bibnamefont {Kawashima}},\ }\bibfield  {title}
  {\enquote {\bibinfo {title} {Possibility of flat-band ferromagnetism in
  hole-doped pyrochlore oxides sn 2 nb 2 o 7 and sn 2 ta 2 o 7},}\ }\href@noop
  {} {\bibfield  {journal} {\bibinfo  {journal} {Physical review letters}\
  }\textbf {\bibinfo {volume} {120}},\ \bibinfo {pages} {196401} (\bibinfo
  {year} {2018})}\BibitemShut {NoStop}%
\bibitem [{\citenamefont {Aoki}\ \emph {et~al.}(1996)\citenamefont {Aoki},
  \citenamefont {Ando},\ and\ \citenamefont {Matsumura}}]{aoki1996hofstadter}%
  \BibitemOpen
  \bibfield  {author} {\bibinfo {author} {\bibfnamefont {Hideo}\ \bibnamefont
  {Aoki}}, \bibinfo {author} {\bibfnamefont {Masato}\ \bibnamefont {Ando}}, \
  and\ \bibinfo {author} {\bibfnamefont {Hajime}\ \bibnamefont {Matsumura}},\
  }\bibfield  {title} {\enquote {\bibinfo {title} {Hofstadter butterflies for
  flat bands},}\ }\href@noop {} {\bibfield  {journal} {\bibinfo  {journal}
  {Physical Review B}\ }\textbf {\bibinfo {volume} {54}},\ \bibinfo {pages}
  {R17296} (\bibinfo {year} {1996})}\BibitemShut {NoStop}%
\bibitem [{\citenamefont {Ramachandran}\ \emph {et~al.}(2017)\citenamefont
  {Ramachandran}, \citenamefont {Andreanov},\ and\ \citenamefont
  {Flach}}]{ramachandran2017chiral}%
  \BibitemOpen
  \bibfield  {author} {\bibinfo {author} {\bibfnamefont {Ajith}\ \bibnamefont
  {Ramachandran}}, \bibinfo {author} {\bibfnamefont {Alexei}\ \bibnamefont
  {Andreanov}}, \ and\ \bibinfo {author} {\bibfnamefont {Sergej}\ \bibnamefont
  {Flach}},\ }\bibfield  {title} {\enquote {\bibinfo {title} {Chiral flat
  bands: Existence, engineering, and stability},}\ }\href@noop {} {\bibfield
  {journal} {\bibinfo  {journal} {Physical Review B}\ }\textbf {\bibinfo
  {volume} {96}},\ \bibinfo {pages} {161104} (\bibinfo {year}
  {2017})}\BibitemShut {NoStop}%
\bibitem [{\citenamefont {Regnault}\ and\ \citenamefont
  {Bernevig}(2011)}]{regnault2011fractional}%
  \BibitemOpen
  \bibfield  {author} {\bibinfo {author} {\bibfnamefont {Nicolas}\ \bibnamefont
  {Regnault}}\ and\ \bibinfo {author} {\bibfnamefont {B~Andrei}\ \bibnamefont
  {Bernevig}},\ }\bibfield  {title} {\enquote {\bibinfo {title} {Fractional
  chern insulator},}\ }\href@noop {} {\bibfield  {journal} {\bibinfo  {journal}
  {Physical Review X}\ }\textbf {\bibinfo {volume} {1}},\ \bibinfo {pages}
  {021014} (\bibinfo {year} {2011})}\BibitemShut {NoStop}%
\bibitem [{\citenamefont {Rhim}\ and\ \citenamefont
  {Yang}(2019)}]{rhim2019classification}%
  \BibitemOpen
  \bibfield  {author} {\bibinfo {author} {\bibfnamefont {Jun-Won}\ \bibnamefont
  {Rhim}}\ and\ \bibinfo {author} {\bibfnamefont {Bohm-Jung}\ \bibnamefont
  {Yang}},\ }\bibfield  {title} {\enquote {\bibinfo {title} {Classification of
  flat bands according to the band-crossing singularity of bloch wave
  functions},}\ }\href@noop {} {\bibfield  {journal} {\bibinfo  {journal}
  {Physical Review B}\ }\textbf {\bibinfo {volume} {99}},\ \bibinfo {pages}
  {045107} (\bibinfo {year} {2019})}\BibitemShut {NoStop}%
\bibitem [{\citenamefont {Ma}\ \emph {et~al.}(2020)\citenamefont {Ma},
  \citenamefont {Rhim}, \citenamefont {Tang}, \citenamefont {Xia},
  \citenamefont {Wang}, \citenamefont {Zheng}, \citenamefont {Xia},
  \citenamefont {Song}, \citenamefont {Hu}, \citenamefont {Li} \emph
  {et~al.}}]{ma2020direct}%
  \BibitemOpen
  \bibfield  {author} {\bibinfo {author} {\bibfnamefont {Jina}\ \bibnamefont
  {Ma}}, \bibinfo {author} {\bibfnamefont {Jun-Won}\ \bibnamefont {Rhim}},
  \bibinfo {author} {\bibfnamefont {Liqin}\ \bibnamefont {Tang}}, \bibinfo
  {author} {\bibfnamefont {Shiqi}\ \bibnamefont {Xia}}, \bibinfo {author}
  {\bibfnamefont {Haiping}\ \bibnamefont {Wang}}, \bibinfo {author}
  {\bibfnamefont {Xiuyan}\ \bibnamefont {Zheng}}, \bibinfo {author}
  {\bibfnamefont {Shiqiang}\ \bibnamefont {Xia}}, \bibinfo {author}
  {\bibfnamefont {Daohong}\ \bibnamefont {Song}}, \bibinfo {author}
  {\bibfnamefont {Yi}~\bibnamefont {Hu}}, \bibinfo {author} {\bibfnamefont
  {Yigang}\ \bibnamefont {Li}},  \emph {et~al.},\ }\bibfield  {title} {\enquote
  {\bibinfo {title} {Direct observation of flatband loop states arising from
  nontrivial real-space topology},}\ }\href@noop {} {\bibfield  {journal}
  {\bibinfo  {journal} {Physical Review Letters}\ }\textbf {\bibinfo {volume}
  {124}},\ \bibinfo {pages} {183901} (\bibinfo {year} {2020})}\BibitemShut
  {NoStop}%
\bibitem [{\citenamefont {Rhim}\ and\ \citenamefont
  {Yang}(2021)}]{rhim2021singular}%
  \BibitemOpen
  \bibfield  {author} {\bibinfo {author} {\bibfnamefont {Jun-Won}\ \bibnamefont
  {Rhim}}\ and\ \bibinfo {author} {\bibfnamefont {Bohm-Jung}\ \bibnamefont
  {Yang}},\ }\bibfield  {title} {\enquote {\bibinfo {title} {Singular flat
  bands},}\ }\href@noop {} {\bibfield  {journal} {\bibinfo  {journal} {Advances
  in Physics: X}\ }\textbf {\bibinfo {volume} {6}},\ \bibinfo {pages} {1901606}
  (\bibinfo {year} {2021})}\BibitemShut {NoStop}%
\bibitem [{\citenamefont {Rhim}\ \emph {et~al.}(2020)\citenamefont {Rhim},
  \citenamefont {Kim},\ and\ \citenamefont {Yang}}]{rhim2020quantum}%
  \BibitemOpen
  \bibfield  {author} {\bibinfo {author} {\bibfnamefont {Jun-Won}\ \bibnamefont
  {Rhim}}, \bibinfo {author} {\bibfnamefont {Kyoo}\ \bibnamefont {Kim}}, \ and\
  \bibinfo {author} {\bibfnamefont {Bohm-Jung}\ \bibnamefont {Yang}},\
  }\bibfield  {title} {\enquote {\bibinfo {title} {Quantum distance and
  anomalous landau levels of flat bands},}\ }\href@noop {} {\bibfield
  {journal} {\bibinfo  {journal} {Nature}\ }\textbf {\bibinfo {volume} {584}},\
  \bibinfo {pages} {59--63} (\bibinfo {year} {2020})}\BibitemShut {NoStop}%
\bibitem [{\citenamefont {Choi}\ \emph {et~al.}(2011)\citenamefont {Choi},
  \citenamefont {Ahn}, \citenamefont {Han}, \citenamefont {Seol}, \citenamefont
  {Moon}, \citenamefont {Kim},\ and\ \citenamefont
  {Choi}}]{choi2011transformable}%
  \BibitemOpen
  \bibfield  {author} {\bibinfo {author} {\bibfnamefont {Sung-Jin}\
  \bibnamefont {Choi}}, \bibinfo {author} {\bibfnamefont {Jae-Hyuk}\
  \bibnamefont {Ahn}}, \bibinfo {author} {\bibfnamefont {Jin-Woo}\ \bibnamefont
  {Han}}, \bibinfo {author} {\bibfnamefont {Myeong-Lok}\ \bibnamefont {Seol}},
  \bibinfo {author} {\bibfnamefont {Dong-Il}\ \bibnamefont {Moon}}, \bibinfo
  {author} {\bibfnamefont {Sungho}\ \bibnamefont {Kim}}, \ and\ \bibinfo
  {author} {\bibfnamefont {Yang-Kyu}\ \bibnamefont {Choi}},\ }\bibfield
  {title} {\enquote {\bibinfo {title} {Transformable functional nanoscale
  building blocks with wafer-scale silicon nanowires},}\ }\href@noop {}
  {\bibfield  {journal} {\bibinfo  {journal} {Nano letters}\ }\textbf {\bibinfo
  {volume} {11}},\ \bibinfo {pages} {854--859} (\bibinfo {year}
  {2011})}\BibitemShut {NoStop}%
\bibitem [{\citenamefont {Jeon}\ \emph {et~al.}(2015)\citenamefont {Jeon},
  \citenamefont {Kim}, \citenamefont {Lim},\ and\ \citenamefont
  {Kim}}]{jeon2015steep}%
  \BibitemOpen
  \bibfield  {author} {\bibinfo {author} {\bibfnamefont {Youngin}\ \bibnamefont
  {Jeon}}, \bibinfo {author} {\bibfnamefont {Minsuk}\ \bibnamefont {Kim}},
  \bibinfo {author} {\bibfnamefont {Doohyeok}\ \bibnamefont {Lim}}, \ and\
  \bibinfo {author} {\bibfnamefont {Sangsig}\ \bibnamefont {Kim}},\ }\bibfield
  {title} {\enquote {\bibinfo {title} {Steep subthreshold swing n-and p-channel
  operation of bendable feedback field-effect transistors with p+--i--n+
  nanowires by dual-top-gate voltage modulation},}\ }\href@noop {} {\bibfield
  {journal} {\bibinfo  {journal} {Nano letters}\ }\textbf {\bibinfo {volume}
  {15}},\ \bibinfo {pages} {4905--4913} (\bibinfo {year} {2015})}\BibitemShut
  {NoStop}%
\bibitem [{\citenamefont {Drozdov}\ \emph {et~al.}(2014)\citenamefont
  {Drozdov}, \citenamefont {Alexandradinata}, \citenamefont {Jeon},
  \citenamefont {Nadj-Perge}, \citenamefont {Ji}, \citenamefont {Cava},
  \citenamefont {Bernevig},\ and\ \citenamefont {Yazdani}}]{drozdov2014one}%
  \BibitemOpen
  \bibfield  {author} {\bibinfo {author} {\bibfnamefont {Ilya~K}\ \bibnamefont
  {Drozdov}}, \bibinfo {author} {\bibfnamefont {Aris}\ \bibnamefont
  {Alexandradinata}}, \bibinfo {author} {\bibfnamefont {Sangjun}\ \bibnamefont
  {Jeon}}, \bibinfo {author} {\bibfnamefont {Stevan}\ \bibnamefont
  {Nadj-Perge}}, \bibinfo {author} {\bibfnamefont {Huiwen}\ \bibnamefont {Ji}},
  \bibinfo {author} {\bibfnamefont {RJ}~\bibnamefont {Cava}}, \bibinfo {author}
  {\bibfnamefont {B~Andrei}\ \bibnamefont {Bernevig}}, \ and\ \bibinfo {author}
  {\bibfnamefont {Ali}\ \bibnamefont {Yazdani}},\ }\bibfield  {title} {\enquote
  {\bibinfo {title} {One-dimensional topological edge states of bismuth
  bilayers},}\ }\href@noop {} {\bibfield  {journal} {\bibinfo  {journal}
  {Nature Physics}\ }\textbf {\bibinfo {volume} {10}},\ \bibinfo {pages}
  {664--669} (\bibinfo {year} {2014})}\BibitemShut {NoStop}%
\bibitem [{\citenamefont {Zhu}\ \emph {et~al.}(2020)\citenamefont {Zhu},
  \citenamefont {Li}, \citenamefont {Yang}, \citenamefont {Nie}, \citenamefont
  {Xu}, \citenamefont {Yang}, \citenamefont {Guan}, \citenamefont {Wang},
  \citenamefont {Li}, \citenamefont {Liu} \emph {et~al.}}]{zhu2020tunable}%
  \BibitemOpen
  \bibfield  {author} {\bibinfo {author} {\bibfnamefont {Zhen}\ \bibnamefont
  {Zhu}}, \bibinfo {author} {\bibfnamefont {Si}~\bibnamefont {Li}}, \bibinfo
  {author} {\bibfnamefont {Meng}\ \bibnamefont {Yang}}, \bibinfo {author}
  {\bibfnamefont {Xiao-Ang}\ \bibnamefont {Nie}}, \bibinfo {author}
  {\bibfnamefont {Hao-Ke}\ \bibnamefont {Xu}}, \bibinfo {author} {\bibfnamefont
  {Xu}~\bibnamefont {Yang}}, \bibinfo {author} {\bibfnamefont {Dan-Dan}\
  \bibnamefont {Guan}}, \bibinfo {author} {\bibfnamefont {Shiyong}\
  \bibnamefont {Wang}}, \bibinfo {author} {\bibfnamefont {Yao-Yi}\ \bibnamefont
  {Li}}, \bibinfo {author} {\bibfnamefont {Canhua}\ \bibnamefont {Liu}},  \emph
  {et~al.},\ }\bibfield  {title} {\enquote {\bibinfo {title} {A tunable and
  unidirectional one-dimensional electronic system nb 2n+ 1 si n te 4n+ 2},}\
  }\href@noop {} {\bibfield  {journal} {\bibinfo  {journal} {npj Quantum
  Materials}\ }\textbf {\bibinfo {volume} {5}},\ \bibinfo {pages} {1--7}
  (\bibinfo {year} {2020})}\BibitemShut {NoStop}%
\bibitem [{\citenamefont {Ahn}\ \emph {et~al.}(2005)\citenamefont {Ahn},
  \citenamefont {Kang}, \citenamefont {Ryang},\ and\ \citenamefont
  {Yeom}}]{ahn2005coexistence}%
  \BibitemOpen
  \bibfield  {author} {\bibinfo {author} {\bibfnamefont {JR}~\bibnamefont
  {Ahn}}, \bibinfo {author} {\bibfnamefont {PG}~\bibnamefont {Kang}}, \bibinfo
  {author} {\bibfnamefont {KD}~\bibnamefont {Ryang}}, \ and\ \bibinfo {author}
  {\bibfnamefont {HW}~\bibnamefont {Yeom}},\ }\bibfield  {title} {\enquote
  {\bibinfo {title} {Coexistence of two different peierls distortions within an
  atomic scale wire: Si (553)-au},}\ }\href@noop {} {\bibfield  {journal}
  {\bibinfo  {journal} {Physical review letters}\ }\textbf {\bibinfo {volume}
  {95}},\ \bibinfo {pages} {196402} (\bibinfo {year} {2005})}\BibitemShut
  {NoStop}%
\bibitem [{\citenamefont {Senkovskiy}\ \emph {et~al.}(2021)\citenamefont
  {Senkovskiy}, \citenamefont {Nenashev}, \citenamefont {Alavi}, \citenamefont
  {Falke}, \citenamefont {Hell}, \citenamefont {Bampoulis}, \citenamefont
  {Rybkovskiy}, \citenamefont {Usachov}, \citenamefont {Fedorov}, \citenamefont
  {Chernov} \emph {et~al.}}]{senkovskiy2021tunneling}%
  \BibitemOpen
  \bibfield  {author} {\bibinfo {author} {\bibfnamefont {Boris~V}\ \bibnamefont
  {Senkovskiy}}, \bibinfo {author} {\bibfnamefont {Alexey~V}\ \bibnamefont
  {Nenashev}}, \bibinfo {author} {\bibfnamefont {Seyed~K}\ \bibnamefont
  {Alavi}}, \bibinfo {author} {\bibfnamefont {Yannic}\ \bibnamefont {Falke}},
  \bibinfo {author} {\bibfnamefont {Martin}\ \bibnamefont {Hell}}, \bibinfo
  {author} {\bibfnamefont {Pantelis}\ \bibnamefont {Bampoulis}}, \bibinfo
  {author} {\bibfnamefont {Dmitry~V}\ \bibnamefont {Rybkovskiy}}, \bibinfo
  {author} {\bibfnamefont {Dmitry~Yu}\ \bibnamefont {Usachov}}, \bibinfo
  {author} {\bibfnamefont {Alexander~V}\ \bibnamefont {Fedorov}}, \bibinfo
  {author} {\bibfnamefont {Alexander~I}\ \bibnamefont {Chernov}},  \emph
  {et~al.},\ }\bibfield  {title} {\enquote {\bibinfo {title} {Tunneling current
  modulation in atomically precise graphene nanoribbon heterojunctions},}\
  }\href@noop {} {\bibfield  {journal} {\bibinfo  {journal} {Nature
  communications}\ }\textbf {\bibinfo {volume} {12}},\ \bibinfo {pages} {1--11}
  (\bibinfo {year} {2021})}\BibitemShut {NoStop}%
\bibitem [{\citenamefont {Karakachian}\ \emph {et~al.}(2020)\citenamefont
  {Karakachian}, \citenamefont {Nguyen}, \citenamefont {Aprojanz},
  \citenamefont {Zakharov}, \citenamefont {Yakimova}, \citenamefont
  {Rosenzweig}, \citenamefont {Polley}, \citenamefont {Balasubramanian},
  \citenamefont {Tegenkamp}, \citenamefont {Power} \emph
  {et~al.}}]{karakachian2020one}%
  \BibitemOpen
  \bibfield  {author} {\bibinfo {author} {\bibfnamefont {Hrag}\ \bibnamefont
  {Karakachian}}, \bibinfo {author} {\bibfnamefont {TT~Nhung}\ \bibnamefont
  {Nguyen}}, \bibinfo {author} {\bibfnamefont {Johannes}\ \bibnamefont
  {Aprojanz}}, \bibinfo {author} {\bibfnamefont {Alexei~A}\ \bibnamefont
  {Zakharov}}, \bibinfo {author} {\bibfnamefont {Rositsa}\ \bibnamefont
  {Yakimova}}, \bibinfo {author} {\bibfnamefont {Philipp}\ \bibnamefont
  {Rosenzweig}}, \bibinfo {author} {\bibfnamefont {Craig~M}\ \bibnamefont
  {Polley}}, \bibinfo {author} {\bibfnamefont {Thiagarajan}\ \bibnamefont
  {Balasubramanian}}, \bibinfo {author} {\bibfnamefont {Christoph}\
  \bibnamefont {Tegenkamp}}, \bibinfo {author} {\bibfnamefont {Stephen~R}\
  \bibnamefont {Power}},  \emph {et~al.},\ }\bibfield  {title} {\enquote
  {\bibinfo {title} {One-dimensional confinement and width-dependent bandgap
  formation in epitaxial graphene nanoribbons},}\ }\href@noop {} {\bibfield
  {journal} {\bibinfo  {journal} {Nature communications}\ }\textbf {\bibinfo
  {volume} {11}},\ \bibinfo {pages} {1--8} (\bibinfo {year}
  {2020})}\BibitemShut {NoStop}%
\bibitem [{\citenamefont {Lin}\ \emph {et~al.}(2018)\citenamefont {Lin},
  \citenamefont {Choi}, \citenamefont {Zhang}, \citenamefont {Qin},
  \citenamefont {Yi}, \citenamefont {Wang}, \citenamefont {Li}, \citenamefont
  {Wang}, \citenamefont {Zhang}, \citenamefont {Sun} \emph
  {et~al.}}]{lin2018flatbands}%
  \BibitemOpen
  \bibfield  {author} {\bibinfo {author} {\bibfnamefont {Zhiyong}\ \bibnamefont
  {Lin}}, \bibinfo {author} {\bibfnamefont {Jin-Ho}\ \bibnamefont {Choi}},
  \bibinfo {author} {\bibfnamefont {Qiang}\ \bibnamefont {Zhang}}, \bibinfo
  {author} {\bibfnamefont {Wei}\ \bibnamefont {Qin}}, \bibinfo {author}
  {\bibfnamefont {Seho}\ \bibnamefont {Yi}}, \bibinfo {author} {\bibfnamefont
  {Pengdong}\ \bibnamefont {Wang}}, \bibinfo {author} {\bibfnamefont {Lin}\
  \bibnamefont {Li}}, \bibinfo {author} {\bibfnamefont {Yifan}\ \bibnamefont
  {Wang}}, \bibinfo {author} {\bibfnamefont {Hui}\ \bibnamefont {Zhang}},
  \bibinfo {author} {\bibfnamefont {Zhe}\ \bibnamefont {Sun}},  \emph
  {et~al.},\ }\bibfield  {title} {\enquote {\bibinfo {title} {Flatbands and
  emergent ferromagnetic ordering in fe$_3$ sn$_2$ kagome lattices},}\
  }\href@noop {} {\bibfield  {journal} {\bibinfo  {journal} {Physical review
  letters}\ }\textbf {\bibinfo {volume} {121}},\ \bibinfo {pages} {096401}
  (\bibinfo {year} {2018})}\BibitemShut {NoStop}%
\bibitem [{\citenamefont {Kang}\ \emph
  {et~al.}(2020{\natexlab{a}})\citenamefont {Kang}, \citenamefont {Ye},
  \citenamefont {Fang}, \citenamefont {You}, \citenamefont {Levitan},
  \citenamefont {Han}, \citenamefont {Facio}, \citenamefont {Jozwiak},
  \citenamefont {Bostwick}, \citenamefont {Rotenberg} \emph
  {et~al.}}]{kang2020dirac}%
  \BibitemOpen
  \bibfield  {author} {\bibinfo {author} {\bibfnamefont {Mingu}\ \bibnamefont
  {Kang}}, \bibinfo {author} {\bibfnamefont {Linda}\ \bibnamefont {Ye}},
  \bibinfo {author} {\bibfnamefont {Shiang}\ \bibnamefont {Fang}}, \bibinfo
  {author} {\bibfnamefont {Jhih-Shih}\ \bibnamefont {You}}, \bibinfo {author}
  {\bibfnamefont {Abe}\ \bibnamefont {Levitan}}, \bibinfo {author}
  {\bibfnamefont {Minyong}\ \bibnamefont {Han}}, \bibinfo {author}
  {\bibfnamefont {Jorge~I}\ \bibnamefont {Facio}}, \bibinfo {author}
  {\bibfnamefont {Chris}\ \bibnamefont {Jozwiak}}, \bibinfo {author}
  {\bibfnamefont {Aaron}\ \bibnamefont {Bostwick}}, \bibinfo {author}
  {\bibfnamefont {Eli}\ \bibnamefont {Rotenberg}},  \emph {et~al.},\ }\bibfield
   {title} {\enquote {\bibinfo {title} {Dirac fermions and flat bands in the
  ideal kagome metal fesn},}\ }\href@noop {} {\bibfield  {journal} {\bibinfo
  {journal} {Nature materials}\ }\textbf {\bibinfo {volume} {19}},\ \bibinfo
  {pages} {163--169} (\bibinfo {year} {2020}{\natexlab{a}})}\BibitemShut
  {NoStop}%
\bibitem [{\citenamefont {Kang}\ \emph
  {et~al.}(2020{\natexlab{b}})\citenamefont {Kang}, \citenamefont {Fang},
  \citenamefont {Ye}, \citenamefont {Po}, \citenamefont {Denlinger},
  \citenamefont {Jozwiak}, \citenamefont {Bostwick}, \citenamefont {Rotenberg},
  \citenamefont {Kaxiras}, \citenamefont {Checkelsky} \emph
  {et~al.}}]{kang2020topological}%
  \BibitemOpen
  \bibfield  {author} {\bibinfo {author} {\bibfnamefont {Mingu}\ \bibnamefont
  {Kang}}, \bibinfo {author} {\bibfnamefont {Shiang}\ \bibnamefont {Fang}},
  \bibinfo {author} {\bibfnamefont {Linda}\ \bibnamefont {Ye}}, \bibinfo
  {author} {\bibfnamefont {Hoi~Chun}\ \bibnamefont {Po}}, \bibinfo {author}
  {\bibfnamefont {Jonathan}\ \bibnamefont {Denlinger}}, \bibinfo {author}
  {\bibfnamefont {Chris}\ \bibnamefont {Jozwiak}}, \bibinfo {author}
  {\bibfnamefont {Aaron}\ \bibnamefont {Bostwick}}, \bibinfo {author}
  {\bibfnamefont {Eli}\ \bibnamefont {Rotenberg}}, \bibinfo {author}
  {\bibfnamefont {Efthimios}\ \bibnamefont {Kaxiras}}, \bibinfo {author}
  {\bibfnamefont {Joseph~G}\ \bibnamefont {Checkelsky}},  \emph {et~al.},\
  }\bibfield  {title} {\enquote {\bibinfo {title} {Topological flat bands in
  frustrated kagome lattice cosn},}\ }\href@noop {} {\bibfield  {journal}
  {\bibinfo  {journal} {Nature communications}\ }\textbf {\bibinfo {volume}
  {11}},\ \bibinfo {pages} {1--9} (\bibinfo {year}
  {2020}{\natexlab{b}})}\BibitemShut {NoStop}%
\bibitem [{\citenamefont {Scott}(2009)}]{scott2000}%
  \BibitemOpen
  \bibfield  {author} {\bibinfo {author} {\bibfnamefont {James~F.}\
  \bibnamefont {Scott}},\ }\bibfield  {title} {\enquote {\bibinfo {title}
  {Ferroelectric memories},}\ }\bibfield  {booktitle} {\emph {\bibinfo
  {booktitle} {Ferroelectric Memories}},\ }\href@noop {} {\  (\bibinfo {year}
  {2009})}\BibitemShut {NoStop}%
\bibitem [{\citenamefont {Ma}\ \emph {et~al.}(2018)\citenamefont {Ma},
  \citenamefont {Ma}, \citenamefont {Zhang}, \citenamefont {Peng},
  \citenamefont {Wang}, \citenamefont {Liu}, \citenamefont {Wang},
  \citenamefont {Li}, \citenamefont {Chen}, \citenamefont {Cheng},
  \citenamefont {Gao}, \citenamefont {Gu}, \citenamefont {Chen}, \citenamefont
  {Yu}, \citenamefont {Zhang},\ and\ \citenamefont {Nan}}]{ma2018controllable}%
  \BibitemOpen
  \bibfield  {author} {\bibinfo {author} {\bibfnamefont {Ji}~\bibnamefont
  {Ma}}, \bibinfo {author} {\bibfnamefont {Jing}\ \bibnamefont {Ma}}, \bibinfo
  {author} {\bibfnamefont {Qinghua}\ \bibnamefont {Zhang}}, \bibinfo {author}
  {\bibfnamefont {Renci}\ \bibnamefont {Peng}}, \bibinfo {author}
  {\bibfnamefont {Jing}\ \bibnamefont {Wang}}, \bibinfo {author} {\bibfnamefont
  {Chen}\ \bibnamefont {Liu}}, \bibinfo {author} {\bibfnamefont {Meng}\
  \bibnamefont {Wang}}, \bibinfo {author} {\bibfnamefont {Ning}\ \bibnamefont
  {Li}}, \bibinfo {author} {\bibfnamefont {Mingfeng}\ \bibnamefont {Chen}},
  \bibinfo {author} {\bibfnamefont {Xiaoxing}\ \bibnamefont {Cheng}}, \bibinfo
  {author} {\bibfnamefont {Peng}\ \bibnamefont {Gao}}, \bibinfo {author}
  {\bibfnamefont {Lin}\ \bibnamefont {Gu}}, \bibinfo {author} {\bibfnamefont
  {Long-Qing}\ \bibnamefont {Chen}}, \bibinfo {author} {\bibfnamefont
  {Pu}~\bibnamefont {Yu}}, \bibinfo {author} {\bibfnamefont {Jinxing}\
  \bibnamefont {Zhang}}, \ and\ \bibinfo {author} {\bibfnamefont {Ce-Wen}\
  \bibnamefont {Nan}},\ }\bibfield  {title} {\enquote {\bibinfo {title}
  {Controllable conductive readout in self-assembled, topologically confined
  ferroelectric domain walls},}\ }\href@noop {} {\bibfield  {journal} {\bibinfo
   {journal} {Nature nanotechnology}\ }\textbf {\bibinfo {volume} {13}},\
  \bibinfo {pages} {947--952} (\bibinfo {year} {2018})}\BibitemShut {NoStop}%
\bibitem [{\citenamefont {Jiang}\ \emph {et~al.}(2018)\citenamefont {Jiang},
  \citenamefont {Bai}, \citenamefont {Chen}, \citenamefont {He}, \citenamefont
  {Zhang}, \citenamefont {Zhang}, \citenamefont {Shi}, \citenamefont {Park},
  \citenamefont {Scott}, \citenamefont {Hwang},\ and\ \citenamefont
  {Jiang}}]{jiang2018temporary}%
  \BibitemOpen
  \bibfield  {author} {\bibinfo {author} {\bibfnamefont {Jun}\ \bibnamefont
  {Jiang}}, \bibinfo {author} {\bibfnamefont {Zi~Long}\ \bibnamefont {Bai}},
  \bibinfo {author} {\bibfnamefont {Zhi~Hui}\ \bibnamefont {Chen}}, \bibinfo
  {author} {\bibfnamefont {Long}\ \bibnamefont {He}}, \bibinfo {author}
  {\bibfnamefont {David~Wei}\ \bibnamefont {Zhang}}, \bibinfo {author}
  {\bibfnamefont {Qing~Hua}\ \bibnamefont {Zhang}}, \bibinfo {author}
  {\bibfnamefont {Jin~An}\ \bibnamefont {Shi}}, \bibinfo {author}
  {\bibfnamefont {Min~Hyuk}\ \bibnamefont {Park}}, \bibinfo {author}
  {\bibfnamefont {James~F}\ \bibnamefont {Scott}}, \bibinfo {author}
  {\bibfnamefont {Cheol~Seong}\ \bibnamefont {Hwang}}, \ and\ \bibinfo {author}
  {\bibfnamefont {An~Quan}\ \bibnamefont {Jiang}},\ }\bibfield  {title}
  {\enquote {\bibinfo {title} {Temporary formation of highly conducting domain
  walls for non-destructive read-out of ferroelectric domain-wall resistance
  switching memories},}\ }\href@noop {} {\bibfield  {journal} {\bibinfo
  {journal} {Nature materials}\ }\textbf {\bibinfo {volume} {17}},\ \bibinfo
  {pages} {49--56} (\bibinfo {year} {2018})}\BibitemShut {NoStop}%
\end{thebibliography}

%

{\small \subsection*{Acknowledgements}
C.-g.O. and J.-W.R. were supported by the National Research Foundation of Korea (NRF) Grant funded by the Korea government (MSIT) (Grant No. 2021R1A2C1010572). J.-W.R. was supported by the National Research Foundation of Korea (NRF) Grant funded by the Korea government (MSIT) (Grant No. 2021R1A5A1032996). D.C. was supported by the National Research Foundation of Korea (NRF) grant funded by the Korea government (MSIT) (Grant No. 2020R1C1C1007895 and 2017R1A5A1014862) and the Yonsei University Research Fund of 2019-22-0209. S.Y.P. was supported by the National Research Foundation of Korea(NRF) grant funded by the Korea government(MSIT) (Grant No. 2021R1C1C1009494) and by Basic Science Research Program through the National Research Foundation of Korea(NRF) funded by the Ministry of Education (Grant No. 2021R1A6A1A03043957).}

{\small \subsection*{Author contributions}
C.-g.O. performed the theoretical analysis, D.C. and S.Y.P. provided experimental perspectives, and J.-W.R. supervised the project. All authors discussed the results and contributed to the final manuscript.
}

{\small \subsection*{Data availability}
The data that support the findings of this study are available from the corresponding authors on reasonable request.
}

{\small \subsection*{Additional information}
Supplementary Information is available in the {\bf online version of the paper.}
Correspondence and requests for materials should be addressed to C.-g.O. or J.-W.R.
}

{\small \subsection*{Competing interest}
The authors declare no competing interest.
}

\newpage

\begin{figure}[t]
\includegraphics[width=85mm]{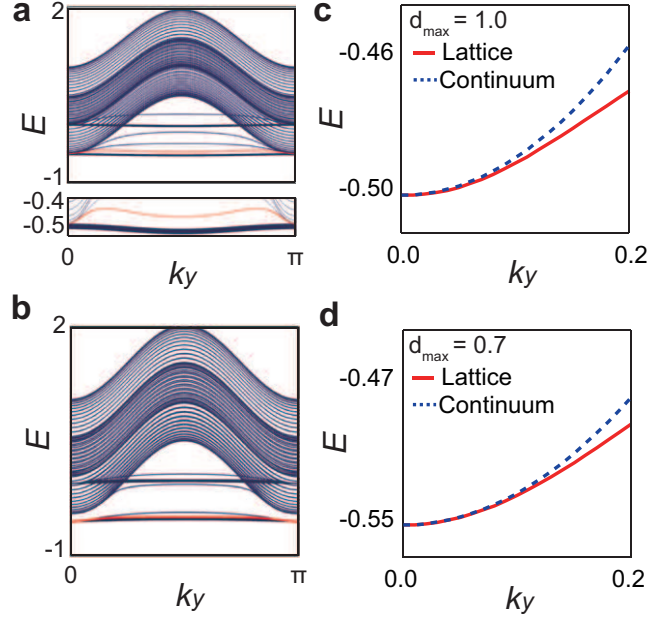} 
\caption{\label{fig5}
\textbf{Interface modes for nearly flat band cases.}
\textbf{a}, The band structure of the square lattice model with perturbation parameters  $(a,b,\Delta)=(-0.01,-0.01,0)$ at $d_{\textrm{max}}=1$. There is a nearly flat band with a parabolic band-crossing at $k_y=0$. In the lower window, the nearly flat band is highlighted.
\textbf{b}, The band dispersion of the square lattice model with perturbation parameters $(a,b,\Delta)=(0,0,0.05)$ at $d_{\mathrm{max}}=0.7$, gaping out the band-touching point and making the flat band nearly flat.
In \textbf{c}(\textbf{d}), we compare the energy of the interface mode in \textbf{a}(\textbf{b}) evaluated from the lattice and continuum model as a function of $k_y$.
}\label{fig:nearly_flat}
\end{figure}

\end{document}